\documentclass{aastex63}

\usepackage{amsmath}
\usepackage{appendix}
\usepackage{xcolor}
\usepackage{float}
\usepackage{diagbox}
\usepackage{bm}
\usepackage{enumitem}
\usepackage{lineno}

\received{January 2024}
\revised{March 2024}
\accepted{March 2024}
\submitjournal{ApJ}

\newcommand{\calbf}[1]{\bm{\mathcal{#1}}}

\begin{document}

\title{ExoCubed: A Riemann-Solver based Cubed-Sphere Dynamic Core for Planetary Atmospheres}

\correspondingauthor{Sihe Chen}
\email{sihechen@caltech.edu }

\author[0000-0002-0901-3428]{Sihe Chen}
\affiliation{Geological and Planetary Sciences, California Institute of Technology}

\author[0000-0002-8280-3119]{Cheng Li}
\affiliation{Climate and Space Sciences and Engineering, University of Michigan}

\begin{abstract}

The computational fluid dynamics on a sphere is relevant to global simulations of geophysical fluid dynamics. Using the conventional spherical-polar (or lat-lon) grid results in a singularity at the poles, with orders of magnitude smaller cell sizes at the poles in comparison to the equator. To address this problem, we developed a general circulation model (dynamic core) with a gnomonic equiangular cubed-sphere configuration. This model is developed based on the Simulating Nonhydrostatic Atmospheres on Planets (SNAP) model, using a finite volume numerical scheme with a Riemann-solver-based dynamic core 
and the vertical implicit correction (VIC) scheme. This change of the horizontal configuration gives a 20-time acceleration of global simulations compared to the lat-lon grid with a similar number of cells at medium resolution. We presented standard tests ranging from 2D shallow-water models to 3D general circulation tests, including earth-like planets and shallow hot Jupiters, to validate the accuracy of the model. The method described in this article is generic to transform any existing finite-volume hydrodynamic model in the Cartesian geometry to the spherical geometry.

\end{abstract}

\section{Introduction}
Circulation alters the thermal structure of the atmosphere and the distribution of chemical tracers within. However, the complex nature of solving partial differential equations (PDE) on a sphere poses a significant challenge to renovating the design of an existing GCM and staying abreast of the progress in numerical techniques.

The diverse climates on exoplanets present both challenges and opportunities for simulating planetary atmospheres. In particular, hot Jupiters exhibit enormous contrasts in temperature between their day and night sides, a phenomenon not observed within our solar system \citep{showman2002atmospheric,Bell2018Increased,Komacek2016Atmospheric}. The large pressure difference may drive supersonic jets and the resulting dissipation could be a significant energy source for the planetary interior \citep{fromang2016shear}. However, Earth-centric GCMs, as well as planetary GCMs that are inherited from Earth GCMs, do not handle the formation and the instability of supersonic jets well: the maximum wind speed on Earth is much less than the sound speed \citep{Nasr_2022}. Moreover, the James Webb Space Telescope (JWST) will reveal the atmospheric composition of mini-Neptunes, which are potentially hydrogen and water-rich planets. Since the mass mixing ratio of water vapor in the atmosphere may be comparable to the hydrogen envelope, the interplay of thermodynamics and atmospheric circulation could be extremely complex. A carefully constructed ``moist" GCM is needed to decipher the future transit observations of mini-Neptunes \citep{constantinou2022characterizing}, suggested by CAMEMBERT, a GCM intercomparison project for mini-neptunes \citep{DuncanArXivCamembert}.

Despite the numerous GCMs developed for studying Earth's atmosphere, with at least 167 different models identified in a recent study of GCM code development \citep{kuma2023climate}, they all come down to about 12 major families, which stem from weather forecasting models funded by meteorological agencies \citep{kuma2023climate}. Due to the sharing of development practices, subroutines, numerical treatments, and parameterization schemes, the biases of the simulations (e.g. climate projection) are statistically correlated \citep{kuma2023climate}. As a result, it is crucially important to have a variety of GCMs available to the community, each constructed with different numerical methods and physical approximations, to allow for a diversity of approaches to solving the unknown.

However, developing a GCM from scratch is a time-consuming task that requires knowledge well beyond a basic understanding of numerical methods and physics. The major difficulty is to deal with the spherical geometry, which is almost unique to geophysical fluid dynamics. 
The spectral method, based on the spherical harmonics, is a popular choice for solving partial differential equations on a sphere that overcomes many of the difficulties associated with the spherical geometry. Yet, the scalability of the spectral method is poor as global communication between computer processors is required.
A grid-based method can achieve high scalability on parallel execution but gridding the sphere introduces many complexities that are not present in the Cartesian geometry.

One way of gridding the sphere is to use unstructured grids, which are both flexible and scalable.
Unstructured grids are widely used in computational fluid dynamics (CFD), especially for simulating flows around complex geometries.
Yet, the performance of unstructured grids is generally inferior to structured grids, especially for problems with simple geometry,
due to the overhead of computing the connectivity between the grid points.
Constructing a GCM on unstructured grids has been attempted since the 1960s
\citep{sadourny1968integration}. However, it did not gain popularity until the environmental model OMEGA was developed in 2000 \citep{Bacon2000OMEGA}, and then THOR GCM
\citep{Mendonça2016THOR, deitrick2020thor2} was developed for planetary atmospheres in 2016. The THOR GCM uses the finite-volume method on icosahedral grids and leverages the GPU computing power to achieve high performance.

\begin{figure}[H]
    \centering
    \includegraphics[width=0.5\textwidth]{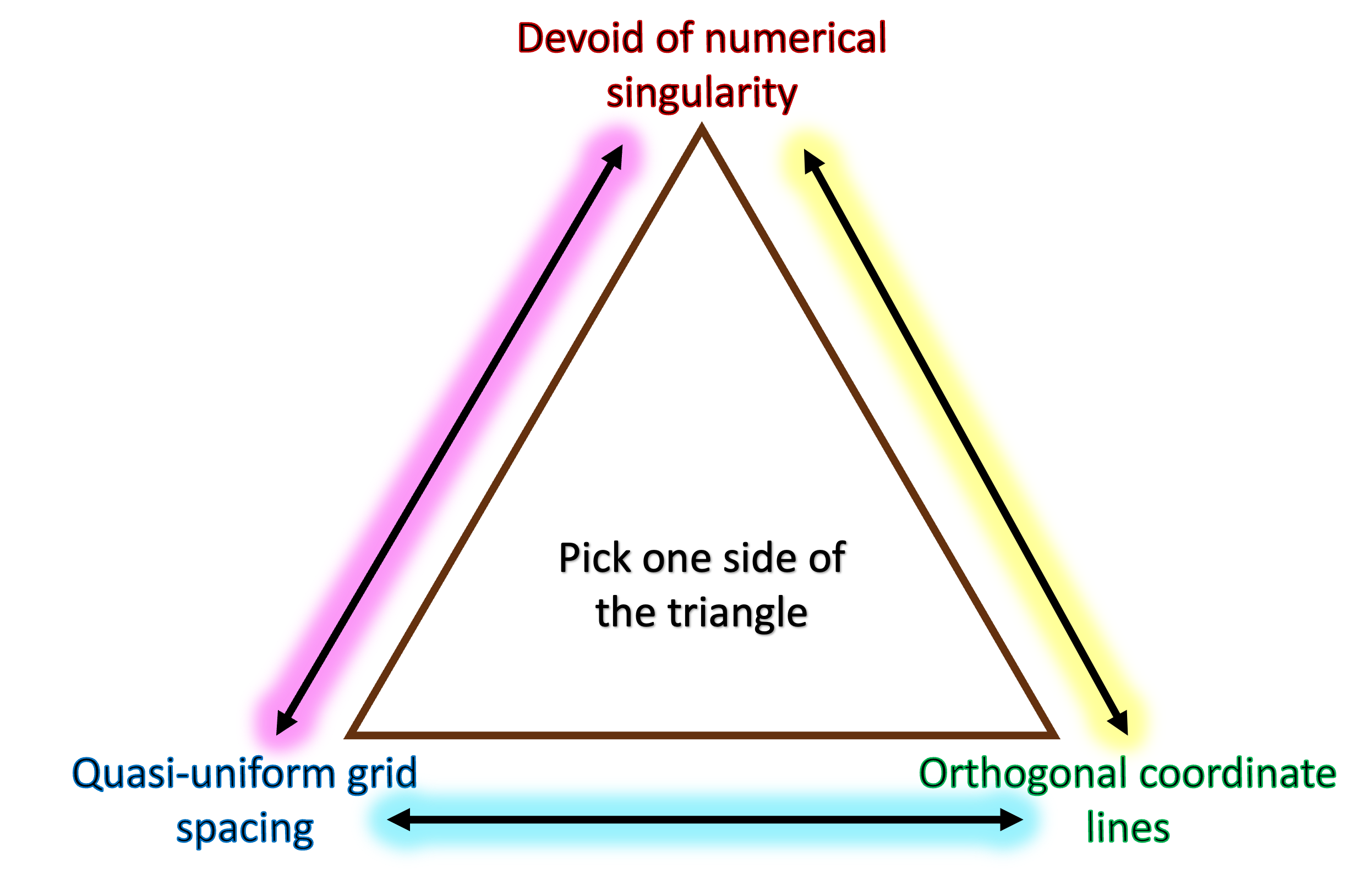}
    \caption{The trilemma for a GCM constructed on a structured gird using explicit time-stepping.}
    \label{fig:trilemma}
\end{figure}

Structured grids can achieve high performance on simple geometry but there is a trilemma for a GCM constructed on a structured grid using explicit time-stepping. That is, we have to sacrifice one of the following three properties: (1) devoid of numerical singularity, (2) quasi-uniform grid spacing, and (3) orthogonal coordinate lines. This is illustrated with Figure \ref{fig:trilemma}.

Here, we regard the numerical singularity as a point where a numerical scheme ceases to be well-behaved. Special treatment is needed to deal with the point at the singularity.
The traditional lat-lon grid has uniform grid spacing and orthogonal coordinate lines, but it has a singularity at each pole.
Due to the converging meridians at the poles, polar filters are needed to stabilize the numerical solution. 
For example, a polar filter is applied to latitudes higher than 45 degrees in the EPIC GCM for planetary atmospheres \citep{Dowling1998epic}
and the Community Atmosphere Model (CAM) for Earth's atmosphere \citep{Neale2010CAM}.
One can avoid the numerical singularity by employing successively coarser grid resolution toward the pole (grid-stretching) at the expense of sacrificing the uniform grid spacing (e.g. \citet{Bindle2021Stretch}).

Mapping a spherical grid to a cube is another way of removing numerical singularity.
The term ``cubed-sphere" refers to a method of tessellating a sphere into six faces, akin to an expanded cube.
The MITgcm \citep{adcroft2004implementation} uses a conformal mapping strategy to
transform the grids on a cube to the sphere while preserving the orthogonality of the coordinate lines.
The conformally mapped cubed-sphere grids have eight singular points, which are the corners of the cube, although they are not as severe as the poles in the lat-lon grid.

Cubed-sphere grids can also be constructed using gnomonic projection, which is non-conformal, but the projected grids are more uniform \citep{chen2021lmars,ullrich2010high}.
Moreover, the gnomonic cubed-sphere grids do not have numerical singularities, i.e. no special treatment is needed at the corners between the six faces.
Yet, it is a non-trivial task to work with the PDE solver on a non-orthogonal geometry. The
Flexible Modeling System (FMS) developed at the Geophysical Fluid Dynamics Laboratory (GFDL) uses the gnomonic cubed-sphere grids for its atmospheric GCM \citep{putman2007finite, mouallem2023implementation}.
Both FMS-based GCMs and the MITgcm have been used to study exoplanet atmospheres \citep{kaspi2015atmospheric, komacek2021constraining}.

In this article, we discuss the development of the dynamic core of a GCM that is designed for general planetary atmospheres that may have shocks, multiple condensing species, or deep atmospheres that are as thick as the planet itself. 
We target the modelers who have access to a hydrodynamic model in Cartesian geometry and we present a general method to convert such existing models to spherical geometry with minimum effort. 
In this way, a variety of hydrodynamic codes can be converted to a GCM, enriching the diversity of GCMs available to the community.
We choose to use the finite volume method on gnomonic cubed-sphere grids as the changes to the existing Cartesian-geometry-solver are minimum among all other choices.

In section 2, we begin with a brief discussion of the finite-volume method and some of its variants.
This section is important for understanding the subsequent choices made in non-orthogonal geometry.
In section 3, we describe the construction of the gnomonic cubed-sphere grids and the computational methods of solving the Euler equations on a cubed-sphere.
In section 4, we present the results of standard test problems and the performance of the cubed-sphere GCM.
In section 5, we summarize and compare our new GCM with other existing ones.

\section{Flavors of the Finite Volume Methods on a Cubed-Sphere}
\label{sec:flavor}
\subsection{Formulation of the equations}
The hydrodynamic equations can be cast into a variety of mathematically equivalent but numerically different forms. We use the momentum equation to elucidate the intricacies of the numerical methods associated with the cubed-sphere geometry, or more generally, with solving the hydrodynamic equations on a sphere. The Eulerian form of the momentum equation is:

\begin{equation}
    \frac{\partial\mathbf{v}}{\partial t} + \mathbf{v}\cdot\nabla\mathbf{v} = - \frac{1}{\rho} \nabla p,
\end{equation}
in which, the $\nabla\mathbf{v}$ is a second-order tensor that is complex to evaluate on a non-orthogonal geometry. It can also be written in the conservative form:
\begin{equation}
    \frac{\partial}{\partial t}\left( \rho \mathbf{v}\right) +\nabla\cdot\left(\rho \mathbf{v}\mathbf{v}+p\mathbf{I}\right)=0
\end{equation}
In the flux-form equations, the conservative variables are used, as opposed to primitive variables. Conservative variables are the quantities that are conserved in a control volume in the absence of sources and sinks, and examples include energy and momentum. Primitive variables are quantities that describe the intensive states of fluid directly, such as temperature and velocities.

One can transform the momentum equation into the vector-invariant form using the following vector identity:

\begin{equation}
    \mathbf{v}\cdot\nabla\mathbf{v} = \frac{1}{2}\nabla |\mathbf{v}|^2 
    - \mathbf{v}\times(\nabla \times \mathbf{v}).
\end{equation}

This formulation eliminates the need for complex metric terms, which typically arise when taking the divergence of a second-order tensor. Moreover, in describing the large-scale circulation of the atmosphere, employing the hydrostatic assumption leads to ``layered atmospheres" that can be solved layer-by-layer using the ``Vertically Lagrangian" approach that treats each layer as a shallow water system \citep{lin2004vertically}. Coupled with equations for tracer advection (e.g. using potential temperature for the energy tracer) on a cubed-sphere \citep{putman2007finite} and a pressure gradient solver \citep{lin1997finite}, an extended equation set based on the shallow water equations, the vertically-Lagrangian advection and conservative remapping scheme became the backbone of the Geophysical Fluid Dynamics Laboratory (GFDL)'s FV3 dynamic core \citep{harris2020nonhydrostatic, mouallem2023implementation}. A similar strategy has been employed by the MITgcm \citep{adcroft2004implementation} where the vector-invariant form is preferred over the conservative momentum equations. Though the MITgcm's cubed-sphere core is restricted to a conformal mapping, and therefore, orthogonal geometry (except at the corners), while the GFDL's cubed-sphere core supports a variety of non-orthogonal mapping choices.

However, the simplicity of avoiding the metric terms comes at the cost of the complex evaluation of the vorticity in a curvilinear geometry. For instance, both the FV3 and MITgcm use the line integral method to assess scalar vorticity. In the case of FV3, velocities are computed on Arakawa-D grids and then interpolated to Arakawa-C grids. Meanwhile, MITgcm computes velocities on Arakawa-C grids and subsequently interpolates them to Arakawa-D grids. The method of calculating the vorticity becomes more intricate when dealing with higher-order methods and three-dimensional non-hydrostatic systems, in which vorticity is a vector rather than a scalar. High-order numerical evaluation of $\mathbf{v}\times(\nabla \times \mathbf{v})$ in a three-dimensional system becomes as complex as computing the metric terms themselves. Thus, it is not obvious that the vector-invariant form of the momentum equation is superior to the Eulerian-from or the conservative flux-form of the momentum equation. 

Solving the conservative flux-form of the shallow water equations has been pioneered by \cite{ullrich2010high}. Let $\sqrt{g} = \big(\det(g_{\mu\nu})\big)^{1/2}$ be the volume element (area element in the two-dimensional sense) of a metric tensor defined by $g_{\mu\nu}$. The flux-form equation reads:
\begin{eqnarray}
    \frac{\partial}{\partial t}\mathbf{Q} + \frac{1}{\sqrt{g}}\frac{\partial}{\partial x^k} \mathbf{F}^k = \bm{\Psi}(\mathbf{Q}) \\
    \mathbf{Q} = (\phi, \phi u^1, \phi u^2)^T \label{eqn:ullrich_q},
\end{eqnarray}
where $\mathbf{F}^k$ is the flux tensor; $\phi$ is the geopotential; $u^1$, $u^2$ are contra-variant velocities; $\bm{\Psi}$ is the forcing term including the pressure gradient, the Coriolis force, and the metric terms. A different choice was made by the seminal work by \cite{sadourny1972conservative}, in which the covariant velocities ($u_1$, $u_2$) were chosen as the prognostic variable. A similar distinction has been made by the FV3 and the MITgcm models, where FV3 uses the covariant velocities as prognostic variables and MITgcm uses the contra-variant velocities as the prognostic variables \citep{harris2020nonhydrostatic,adcroft2004overview}. This deliberate choice hasn't been extensively detailed in official publications. So, we refrain from speculating on the authors' underlying rationale and instead present our logic here.

In a curvilinear coordinate system, contra-variant velocities are tangential to their respective coordinate lines, while covariant velocities are perpendicular to the other coordinate lines. In an orthogonal coordinate system, such as lat-lon or conformal mapping coordinates, both contra-variant and covariant velocities align in the same direction. However, in non-orthogonal systems like the gnomonic equiangle coordinate system, they are different in direction. For any field (scalar or vector), the transport velocity is dictated by the contra-variant velocity, but the off-diagonal terms of the stress tensor will only vanish if it is of the mixed type (having both upper and lower indices). Let $T$ be the stress tensor in the Euler equations:
\begin{equation}
    T^{\mu\nu} = \rho v^\mu v^\nu + g^{\mu\nu}p.
\end{equation}
For a non-orthogonal coordinate system, the metric tensor $g^{\mu\nu}$ has off-diagonal terms, meaning that the pressure gradient in one direction forces the contra-variant velocity in another direction as well. Yet, the symmetric stress tensor reads as:
\begin{equation}
    T^\nu_\mu = \rho v^\nu v_\mu + \delta^\nu_\mu p,
\end{equation}
where $\delta^\nu_\mu$ is the Kronecker delta, which ensures this term appears only in the diagonal entries of the 2D tensor. This ensures that the pressure gradient in one direction only contributes to the covariant velocity tendency in the same direction regardless of the orthogonality of the coordinate system. In general relativistic magnetohydrodynamics, a mixed form of contra-variant and covariant indices for the stress tensor ($T_\mu^\nu$) is also favored because for matrics with an ignorable coordinate $X^\mu$, the flux divergence and the metric terms will vanish for $T_\mu^\nu$ but not for $T^{\mu\nu}$ \citep{gammie2003harm}. As a result, we will use the covariant momentum $(\rho v_\mu)$ as the prognostic variable and diagnose the contra-variant velocity $(v^\nu)$ for transport calculation. The relation between the covariant and contra-variant velocity is $v_\mu = g_{\mu\nu} v^\nu$. In this article, the Einstein summation convention is used, where an index variable appearing both in the upper and lower indices implies summation over all values of the index variable.

\subsection{Choice of basis vectors}
The choice of basis vector is another often overlooked feature in formulating a cubed-sphere GCM. In differential geometry, the coordinate basis is a vector describing the change of position $\mathbf{P}$ with respect to the change of an underlying coordinate variable $X$, i.e., $\partial{\mathbf{P}}/\partial{X}$. In the gnomonic equiangle coordinate system, the coordinate variables for each panel are the angles ($\xi,\eta$) within the range $(-\pi/4,\pi/4)$. The coordinate basis is, thus, not normal and the corresponding velocity is the angular velocity rather than the linear velocity. Thus, there is a choice to be made between the coordinate basis and the normalized non-coordinate basis.

With the coordinate basis, the connection coefficients, or the Christoffel symbols, $\Gamma^\gamma_{\alpha\beta}$, are symmetric in the lower indices $\alpha$ and $\beta$. Moreover, the metric term does not depend on the pressure as $\Gamma^\gamma_{\alpha\beta} g^{\alpha\beta}=0$, which improves the accuracy of the model. \cite{ullrich2010high} solved the shallow water equations using the coordinate basis of the gnomonic equiangle coordinate system. Using the normalized non-coordinate basis results in the asymmetric connection coefficients and the emergence of pressure in the metric terms, as is the case with solving the Euler equations in the traditional lat-lon geometry. To achieve a stable solution, the numerical evaluation of the metric terms should be carefully examined. \cite{chen2021lmars} utilized normalized non-coordinate basis, but avoided calculating the metric terms by solving the vector-invariant form of the shallow water equations.

On the other hand, using the normalized non-coordinate basis leads to a simpler metric tensor and improves the geometric calculation and interpretation. For example, the two-dimensional metric tensor of a gnomonic equiangle coordinate system using the coordinate basis (e.g. \citet{ullrich2010high,ronchi1996cubed,nair2005discontinuous}) is:
\begin{equation}
    g_{\mu\nu} = \frac{(r)^2\big(1+(x)^2)(1+(y)^2\big)}{(\delta)^4}
    \begin{pmatrix}
        1 + (x)^2 & - xy \\
        -xy & 1+ (y)^2
    \end{pmatrix}, \label{eqn:gij1}
\end{equation}
with
\begin{eqnarray}
    x & = & \tan{\xi} \\
    y & = & \tan{\eta} \\
    \delta & = & \sqrt{1 + (x)^2 + (y)^2}.
\end{eqnarray}
Here we use $(\;)^a$ to denote a power law expression rather than a superscript expression in tensor notation. 

If we use normalized non-coordinate basis \citep{chen2021lmars}, the two-dimensional metric tensor of a gnomonic equiangle coordinate system takes a simpler form:
\begin{equation}
    g_{\mu\nu} = 
    \begin{pmatrix}
        1 & \cos\Theta \\
        \cos\Theta & 1
    \end{pmatrix}, \label{eqn:gij2}
\end{equation}
where
\begin{eqnarray}
    \cos\Theta & = & -\frac{xy}{CD} \\
    C & = & \sqrt{1 + (x)^2} \\
    D & = & \sqrt{1 + (y)^2}.
\end{eqnarray} \label{eqn:cosCD}
Equation (\ref{eqn:gij2}) provides a clear geometric interpretation, suggesting that the coordinate lines form an angle of $\Theta$. This is more intuitive than what's presented in equation (\ref{eqn:gij1}). Additionally, equation (\ref{eqn:gij2}) demands fewer float-point operations, consumes less memory, and thus enhances the model's performance. Since the basis vector is normalized, it is the linear velocity that is solved as the prognostic variable. Area and volume preserve their normal geometric meaning. As a consequence, the numerical solver is almost exactly the same as that in the Cartesian geometry except for the non-orthogonal component indicated by $\Theta$. Since the cross-term in the pressure gradient is eliminated by using the covariant momentum, the only extra step in the non-orthogonal cubed-sphere solver compared to a Cartesian geometry solver is to project the flux to the normal direction of the interface between two finite-volume cells. This provides the possibility of adapting many existing Cartesian-geometry-based models into a spherical GCM.

In summary, no single formulation stands out as the universal superior. Each formulation has its own flavor, and every choice inevitably impacts subsequent decisions, e.g. the decision to avoid the metric terms leads to the vector-invariant form of the momentum equations; the need to accurately calculate the vorticity in the vector-invariant form leads to advancing the velocities on Arakawa-D grids and interpolating to Arakawa-C grids. In this paper, the Arakawa-A grid is used with a normalized non-coordinate basis. 

\subsection{Objectives of this work}
The goal pursued in this work is to offer the pathway of adapting models designed for lat-lon or Cartesian grids to the cubed-sphere grid. We describe the minimal recipes needed to facilitate such transformation. This approach effectively bridges the gap between small-scale regional simulations, typically conducted on Cartesian grids, and large-scale simulations that must accommodate a planet's spherical shape. Hence, all thermodynamics and microphysical modules from small-scale models can be seamlessly integrated into large-scale frameworks, eliminating the need for redundant developments.

Our base model to build the cubed-sphere dynamic core is the SNAP model \citep{Li20219SNAP}, which extends the Athena++ general relativistic magnetohydrodynamics code \citep{stone2020athena++} for atmospheric simulation.
We deliberately made subroutines for the cubed-sphere geometry non-intrusive to the original SNAP code, meaning that our model passes all original Athena++ tests and the tests published in the SNAP model. Therefore, our dynamic core, named ExoCubed, serves as a standalone set of subroutines and methods that can be incorporated into any hydrodynamic code as long as the code uses the finite volume method to solve hydrodynamics. 

Particularly, the ExoCubed dynamic core assumes that the base hydrodynamic model has a partial or full implementation of the following features:

\begin{enumerate}
    \item \textbf{Three-dimensional Structured Mesh.} The base model incorporates a mesh in three spatial dimensions, similar to a Cartesian geometry.
    \item \textbf{Dimensional Configuration.} The first dimension is the radial (vertical) direction, denoted as $X^1$. The remaining two dimensions, $X^2$ and $X^3$, are two horizontal dimensions. An illustration of the geometries of the dimensions can be found in Figure \ref{fig:geometry}. This configuration is compatible with a hydrodynamic model using the spherical polar coordinate system but not particularly compatible with the spherical lat-lon geometry, where the third dimension is typically vertical.
    \item \textbf{Grid Placement.} Hydrodynamic variables are placed on collocated grids, also known as the Arakawa-A grid. 
    \item \textbf{Reconstruction.} The left/right states at the cell interface are constructed via a high-order monotonic preserving scheme.
    \item \textbf{Riemann Solver.} First introduced by \citet{Godunov1959Finite}, Riemann Solvers have become widely used in finite volume methods for compressible flows. The Riemann Solver gives a solution to the Riemann problem, which is an initial value problem that is piecewise constant at either side of the interface face. The shocks and rarefaction waves appear as characteristics of the solution. A Riemann Solver is applied in this model to calculate the flux given the left/right hydrodynamic states.
    \item \textbf{Vertically Implicit.} A 1D (vertical) implicit correction/integration scheme is employed to overcome the problem of large aspect ratios. 
    \item \textbf{Dual Variables.} The dynamic equations are evolved using primitive/conserved variable pairs.
    \item \textbf{Variable Configuration.} The primitive variable uses contra-variant velocity and the conserved variable uses covariant momentum.
\end{enumerate}

The more features the base model supports, the easier it will be to adapt to the cube-sphere geometry. In the context of the finite-volume numerical method, as discussed in \cite{leveque2002finite}, a standard high-velocity, compressible hydrodynamic code in the Cartesian geometry generally satisfies all features.
Then, extending the base model to the cubed-sphere geometry requires the following extra steps:

\begin{enumerate} [resume]
    \item \textbf{Mesh on Six Panels.} The spatial dimension is divided into six panels, with the computational domain divided into two panels along one axis and three panels along another. The limits of the spatial dimension are arbitrarily defined in this case. The coordinate lines and grids are uniformly distributed within a panel but are discontinuous across panel boundaries. Such design is to fit the Cubed-Sphere geometry into the Athena++ mesh and parallelization framework. \label{item:first}
    
    \item \textbf{Topological connection.} Since the spherical shell is divided into 6 panels, the topological connections between the panels are different from their relative placement in the coordinate axes. The intra-panel communication is consistent with that in the Cartesian geometry, but the inter-panel communication needs to be specially dealt with by rotation and interpolation (Figure \ref{fig:geometry}).
    
    \item \textbf{Velocity rotation and interpolation.} Ghost zones are layers of extra computational cells outside of block boundaries used for reconstruction but not updated by this block. The ghost zones of one panel cross the panel boundary, and the velocities defined on the neighboring panel need to be rotated and interpolated to the ghost zones of the current panel.

    \item \textbf{Left/right states synchronization.} At the interface between two panels, the left/right states reconstructed by the neighboring panels may differ due to the velocity rotation and interpolation. We devised a synchronization step that uses the right state reconstructed by the right panel and the left state reconstructed by the left panel to be the left/right states at the panel interface to ensure that the flux calculation is consistent across panel boundaries.
    
    \item \textbf{Projection and de-projection.} The dynamic core is designed to be compatible with any existing Riemann Solver. So, we project the left/right states to the local orthonormal coordinate system defined by the cell interface and use the existing Riemann Solver to solve for the fluxes across the interface. Then, we de-project the fluxes to the original coordinate system on each panel (Figure \ref{fig:rsolver}).
    
    \item \textbf{Metric Terms.} We implement the metric terms as the source terms for the momentum equation. We make sure that no spurious flow is generated for a uniform pressure and density field on a sphere. \label{item:last}
\end{enumerate}

In the next section, we discuss in detail the theory of solving the conservative form of the compressible Euler equations on a gnomonic equiangular cubed-sphere geometry defined by normalized (units) basis vectors. We elaborate on
steps \ref{item:first} $\sim$ \ref{item:last}, beginning with an introduction to the gnomonic equiangular coordinate system and ending with the metric terms.

\section{The ExoCubed dynamic core}

\subsection{The gnomonic equiangular coordinate system} \label{section:coordinate}
The grids of the cubed-sphere geometry are obtained by projecting the rectangular grids on the surface of a reference cube radially onto the surface of a concentric sphere \citep{sadourny1972conservative,ronchi1996cubed}, known as the gnomonic projection. There are a few options for choosing the discrete grids on the reference cube. One of them is called the equiangular grids, where the face of the reference cubed is tessellated into $N^2$ number of rectangular grids defined by two angular coordinates $\{\xi,\eta\}$ \citep{ronchi1996cubed}. Suppose that the reference cube is bounded by $[-1,1]\times[-1,1]\times[-1,1]$ concentric with the sphere, $\{\xi,\eta\}$ are uniform distributed angular variables to span the range $(-\pi/4,\pi/4)$, such that:
\begin{eqnarray}
    \xi(j) &=& -\frac{\pi}{4}+(j-\frac{1}{2})\frac{\pi}{2N} \\
    \eta(k) &=& -\frac{\pi}{4}+(k-\frac{1}{2})\frac{\pi}{2N},
\end{eqnarray}
where $(j,k)$ are both integer numbers from $1$ to $N$ to denote the index of a finite volume cell. Thus, the horizontal span of a finite volume cell is bounded by $[\xi(j-1/2),\xi(j+1/2)]\times[\eta(k-1/2),\eta(k+1/2)]$ with the center at $(\xi(j),\eta(k))$. The tangent values of the angular variables $\{\xi,\eta\}$ define the local Cartesian coordinates of the grid points on the surface of the reference cube:
\begin{equation}
    x=\tan\xi,\quad y=\tan\eta, \label{eqn:xy}
\end{equation}
with the derivatives satisfying
\begin{equation}
    \frac{dx}{d\xi}=1+(x)^2,\quad \frac{dy}{d\eta}=1+(y)^2.
\end{equation}

The coordinates described above are defined locally at each face of the reference cube. Therefore, there are six local Cartesian coordinate systems. We give numbers and names to those faces and define a global Cartesian coordinate system that aligns with the local coordinates of the face at the top. Let face~1 be the face at the top, face~5 be the opposite face at the bottom and faces~2,3,4,6 be the faces at four sides. Let the $z$-axis point toward the top, $x$-axis toward the front and $y$-axis toward the right, the local to global transformation of the Cartesian coordinates can be obtained by vector rotation and is summarized in Table~\ref{tab:transform}. Note that we have used $(z,x,y)$ ordering rather than the traditional $(x,y,z)$ ordering of the coordinates due to the convention that the first axis is vertical (radial).

\begin{table}[H]
\centering
\caption{Local to global coordinate system transformation}
\label{tab:transform}
\begin{tabular}{ccc}
\hline
 Face number & Face name & Local to global transformation \\ 
\hline
 1 & top & $(z,x,y) \rightarrow (z,x,y)$ \\ 
 2 & front & $(z,x,y) \rightarrow (-x,z,y)$ \\ 
 3 & left & $(z,x,y) \rightarrow (-z,-x,y)$ \\ 
 4 & right & $(z,x,y) \rightarrow (-x,-y,z)$ \\ 
 5 & bottom & $(z,x,y) \rightarrow (-x,y,-z)$ \\
 6 & back & $(z,x,y) \rightarrow (-x,-z,-y)$ \\ \hline
\end{tabular}
\end{table}

\begin{figure}[H]
    \centering
    \includegraphics[width=0.8\textwidth]{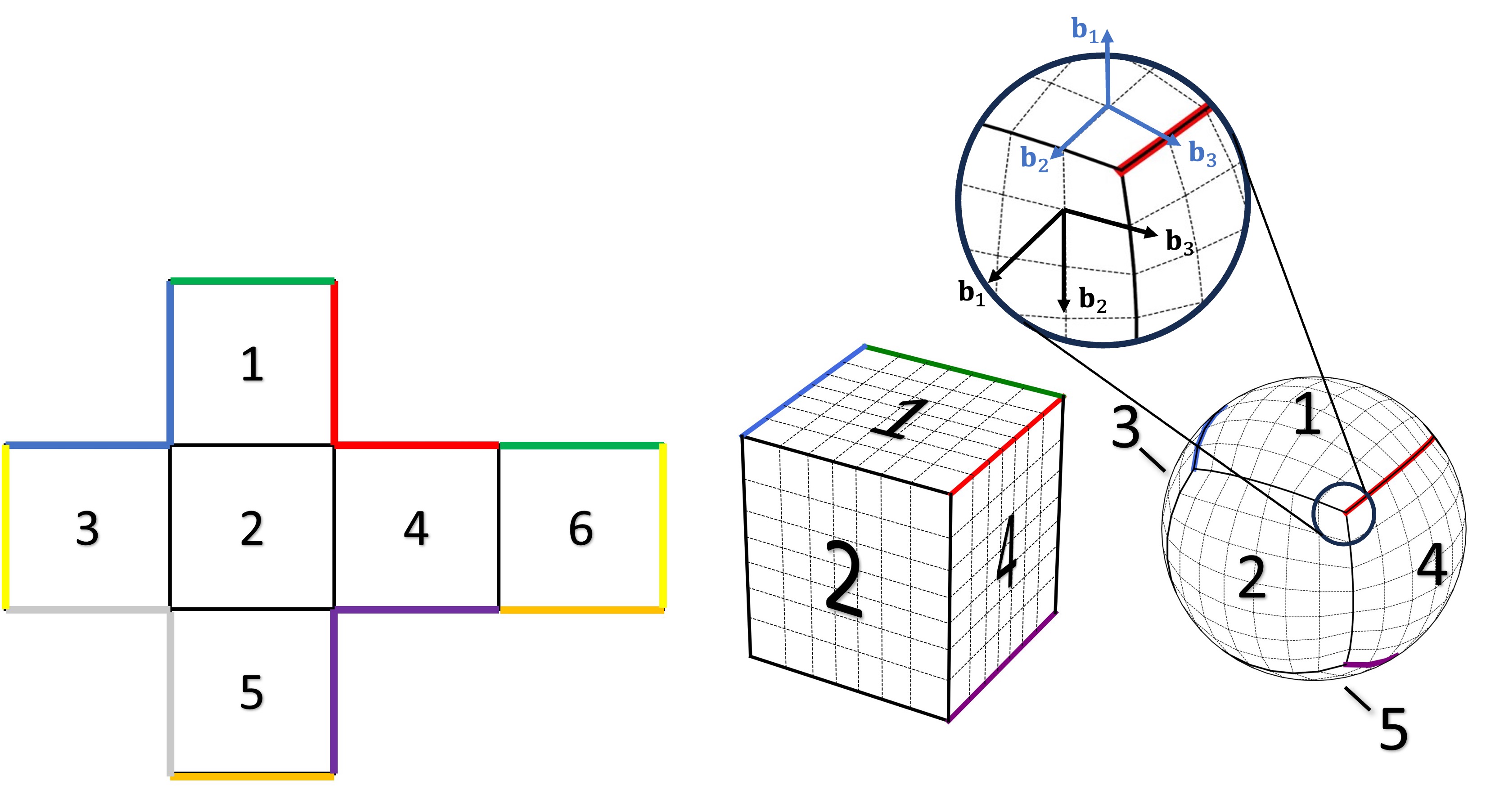}
    \caption{The flattened geometry (left), the folded cube (middle), and the cubed sphere geometry (right) of the six panels of the cubed-sphere. The outer panel boundaries noted with the same color are in touch. In the enlarged view, the coordinate basis of panel 1 (blue) and panel 2 (black) are shown.}
    \label{fig:geometry}
\end{figure}

Projecting the local rectangular grids $(x,y)$ on each face of the reference cube to a concentric sphere results in a curvilinear coordinate system defined on each panel of the cubed sphere. Each point on a sphere is thus described by the radial distance to the center $r$, two angular coordinates $\{\xi,\eta\}$, and the panel (face) number~$n$. Similar to the notation used in \cite{ronchi1996cubed}, we define the following auxiliary pre-computed variables for calculating the geometric relations:
\begin{eqnarray}
    \delta & = & \sqrt{1 + (x)^2 + (y)^2} \label{eqn:delta} \\
    C & = & \sqrt{1 + (x)^2} \\
    D & = & \sqrt{1 + (y)^2},
\end{eqnarray}
satisfying $(\delta)^2+(xy)^2=(CD)^2$.

We present the calculation of the geometric relation of panel~1, the top panel containing the north pole. The geometric relation of the other panels can be derived similarly by rotating the coordinate axis. The starting point of constructing the cubed-sphere coordinate system is the position vector of a point on a spherical panel, whose Cartesian components are:
\begin{equation}
    \mathbf{r}=\frac{r}{\delta}(1,x,y). \label{eqn:rvec}
\end{equation}
The expression of the position vector can be similarly obtained for an oblate spheroid. In problems of rapidly rotating giant planets, the projection of an oblate spheroid becomes a more proper representation. We leave the discussion of the oblate spheroid in Appendix \ref{sec:oblate} and continue with the discussion of the spherical case in the main article.

From the position vector, we obtain the coordinate basis vectors $\{\mathbf{z}_1,\mathbf{z}_2,\mathbf{z}_3\}$ by differentiating the position vector with respect to each coordinate variable:
\begin{equation}
    \mathbf{z}_1=\frac{\partial\mathbf{r}}{\partial r}, \quad
    \mathbf{z}_2=\frac{\partial\mathbf{r}}{\partial \xi}, \quad
    \mathbf{z}_3=\frac{\partial\mathbf{r}}{\partial \eta}.
    \label{eqn:nonbasis}
\end{equation}

As discussed in Section~\ref{sec:flavor}, we favor basis vectors that are of unit length to be compatible with the existing Cartesian model. We define the lengths of the coordinate basis vectors as:
\begin{equation}
    Z_1=|\mathbf{z}_1|=1, \quad
    Z_2=|\mathbf{z}_2|=rD\Big(\frac{C}{\delta}\Big)^2, \quad
    Z_3=|\mathbf{z}_3|=rC\Big(\frac{D}{\delta}\Big)^2.
\end{equation}
Thus, the normalized non-coordinate unit basis vectors are:
\begin{eqnarray}
    \mathbf{b}_1 & = & \frac{\mathbf{z}_1}{Z_1}
        = (\frac{1}{\delta},\frac{x}{\delta},\frac{y}{\delta}) \label{eqn:b1} \\
    \mathbf{b}_2 & = & \frac{\mathbf{z}_2}{Z_2}
        = \frac{1}{D\delta}(-x,1+(y)^2,-xy) \\
    \mathbf{b}_3 & = & \frac{\mathbf{z}_3}{Z_3}
        = \frac{1}{C\delta}(-y,-xy,1+(x)^2) \label{eqn:b3},
\end{eqnarray}
Unit basis vectors $\{\mathbf{b}_1,\mathbf{b}_2,\mathbf{b}_3\}$ constitute the non-coordinate basis at each point on the spherical panel. On each panel, $\mathbf{b}_1$ points at the radial direction normal to $\mathbf{b}_2$ and $\mathbf{b}_3$, but $\mathbf{b}_2$ and $\mathbf{b}_3$ are at an angle $\Theta$ to each other, which is:
\begin{eqnarray}
    \cos\Theta &=& \mathbf{b}_2\cdot\mathbf{b}_3
        = - \frac{xy}{CD} \\
    \sin\Theta &=& \frac{\delta}{CD}.
\end{eqnarray}

Now we switch to the generic tensor notation to calculate the metric and the connection coefficients defined by the unit basis vectors. Let the generic coordinate variables $\{X^1,X^2,X^3\}$ be:
\begin{equation}
    X^1=r, \quad
    X^2=\xi, \quad
    X^3=\eta. \label{eqn:coord}
\end{equation}
The covariant metric tensor is given by the pairwise inner products of the basis vectors:
\begin{equation}
    g_{\mu\nu} = \textbf{b}_\mu \cdot \textbf{b}_\nu = \begin{bmatrix}
1 & 0 & 0\\
0 & 1 & \cos\Theta \\
0 & \cos\Theta & 1
\end{bmatrix},
\end{equation}
and the contra-variant metric tensor $g^{\mu\nu}$ is the inverse of it:
\begin{equation}
g^{\mu\nu} = (g_{\mu\nu})^{-1} = \begin{bmatrix}
1 & 0 & 0\\
0 & \sin ^{-2} \Theta & -\cos\Theta \sin^{-2}\Theta \\
0 & -\cos\Theta \sin^{-2}\Theta & \sin ^{-2} \Theta
\end{bmatrix}.
\label{eq:mettensor}
\end{equation}
We also define
\begin{equation}
    g = |g_{\mu\nu}| = \Big(\frac{\delta}{CD}\Big)^2
\end{equation}
as the determinant of the covariance metric tensor and its square root, $\sqrt{g}=\delta/(CD)=\sin\Theta$, the volume element. 

According to the definition of the basis vectors in equations (\ref{eqn:b1}) - (\ref{eqn:b3}), the connection coefficients of the unit basis (Christoffel symbols) are:
\begin{equation}
    \Gamma^\alpha_{\beta\gamma} = \mathbf{b}^\alpha\cdot\frac{1}{Z_\gamma}
    \frac{\partial \mathbf{b}_\beta}{\partial X^\gamma},
\end{equation}
with the non-zero terms being:
\begin{eqnarray}
    \Gamma^1_{22} & = & \Gamma^1_{33} = -\frac{1}{r}, \quad
    \Gamma^1_{23} = \Gamma^1_{32} = \frac{xy}{rCD} \label{eqn:gamma1} \\
    \Gamma^2_{12} & = & \frac{1}{r}, \quad
    \Gamma^2_{23} = -\frac{y}{rC}\Big(\frac{x}{D}\Big)^2, \quad
    \Gamma^2_{32} = -\frac{y}{rC} \\
    \Gamma^3_{13} & = & \frac{1}{r}, \quad
    \Gamma^3_{23} = -\frac{x}{rD}, \quad
    \Gamma^3_{32} = -\frac{x}{rD}\Big(\frac{y}{C}\Big)^2 \label{eqn:gamma2}
\end{eqnarray}

In a perfect spherical geometry, the radial direction aligns with the vertical direction opposite to the gravitational acceleration vector. So we will use the radial direction and vertical direction interchangeably in this article. We also define the $X^1$-face as the two-dimensional surface that lies on a constant value of $X^1$ and likewise for the $X^2$-face and the $X^3$-face. 

\begin{figure} [H]
    \centering
    \includegraphics[width=0.7\textwidth]{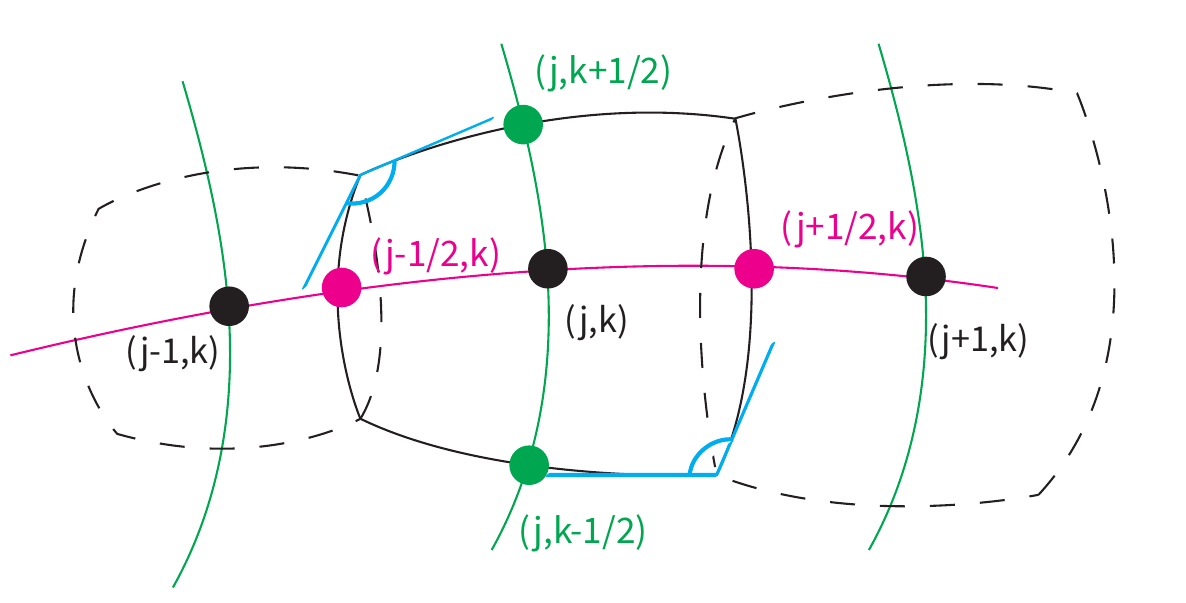}
    \caption{Illustration of non-orthogonal curvilinear geometry. Cell centers are represented by black dots, while face centers by colored dots. Cell centers are marked with integer indices, while cell faces by half-integer indices. Curvilinear coordinate lines are presented in pink and green colors, with the non-orthogonal angle highlighted in blue.}
    \label{fig:cubed-grid}
\end{figure}

\subsection{Equation of motion in a curvilinear geometry}
We solve the standard compressible Euler equations in the curvilinear geometry on a rotating sphere under a potential field. The conservative form of the momentum equation reads as:
\begin{equation}
    \frac{\partial}{\partial t}(\rho\mathbf{v})+\nabla \cdot (\rho \mathbf{v}\mathbf{v})+2 \mathbf{\Omega}\times \rho \textbf{v}=-\nabla p + \rho\nabla\phi,
    \label{eq:mom}
\end{equation}
where $\rho$ is the density, $\mathbf{v}$ the velocity vector, $\mathbf{\Omega}$ the angular velocity of the planet rotation, $p$ the pressure and $\phi$ the potential field.
Expanding the momentum equation in the vector form Equation \eqref{eq:mom} to the component form defined by 
the covariant basis, we have the following form after the arrangement:
\begin{equation}
    \frac{\partial}{\partial t}(\rho v_\alpha)+\nabla_\beta(\rho v_\alpha v^\beta)+2\rho\varepsilon_{\alpha\beta\gamma}\Omega^\beta v^\gamma=-\nabla_\alpha p + \rho\nabla\phi
\end{equation}
Here $\varepsilon_{\alpha\beta\gamma}$ is the Levi-Civita symbol, which is used to write the tensor notation of a cross product: $\mathbf{U}\times\mathbf{V}=\varepsilon_{ijk}U^iV^j\hat{\mathbf{e}}^k$ (here $\mathbf{e}^k$ is the basis vector in $k$-direction). The divergence theorem on the second-order tensor states that \citep{Grinfeld2014Introduction}:
\begin{equation}
    \nabla_\beta T_\alpha^\beta=\sum_\beta\frac{1}{Z_\beta}
    \frac{\partial T_\alpha^\beta}{\partial X^\beta}
    +\Gamma^\gamma_{\gamma\beta}T_\alpha^\beta
    -\Gamma^\gamma_{\alpha\beta}T_\gamma^\beta
    =\sum_\beta\frac{1}{\sqrt{g}Z_\beta}\frac{\partial}{\partial X^\beta}(\sqrt{g}T_\alpha^\beta)-\Gamma^\gamma_{\alpha\beta}T_\gamma^\beta
\end{equation}
Thus the momentum equations in the covariant basis are written as:
\begin{equation}
    \frac{\partial}{\partial t}(\rho v_\alpha)
    + \sum_\beta\frac{1}{\sqrt{g}Z_\beta}
    \frac{\partial}{\partial X^\beta}(\sqrt{g}T_\alpha^\beta)=
    -2\varepsilon_{\alpha\beta\gamma}\Omega^\beta\rho v^\gamma
    +\Gamma^\gamma_{\alpha\beta}T_\gamma^\beta \label{eqn:rhov_t}
    +\rho\nabla_\alpha\phi,
\end{equation}
where 
\begin{equation}
    T_\alpha^\beta=\rho v_\alpha v^\beta+\delta_\alpha^\beta p, \label{eqn:tab}
\end{equation}
is the total stress tensor and the terms on the right-hand-side of the equation above are the Coriolis force, the metric term and the gravitational force, respectively. Note that pressure only exists in the diagonal terms due to the choice of mixed lower and upper indices (See the discussion in Section \ref{sec:flavor}). 

Similarly, the continuity and energy equations without external forcing are:
\begin{equation}
    \frac{\partial}{\partial t}\rho + \sum_\beta\frac{1}{\sqrt{g}Z_\beta}\frac{\partial}{\partial X^\beta}(\sqrt{g}\rho v^\beta)=0 \label{eqn:rho}
\end{equation}
\begin{equation}
    \frac{\partial}{\partial t}(\rho e) + \sum_\beta\frac{1}{\sqrt{g}Z_\beta}\frac{\partial}{\partial X^\beta}[\sqrt{g}(\rho e+p) v^\beta]=\rho v^\beta\nabla_\beta\phi, \label{eqn:rhoe}
\end{equation}
where $e$ is the sum of specific internal energy and kinetic energy. For example, for an ideal gas, it is defined as:
\begin{equation}
    e=\frac{1}{\gamma-1}\frac{p}{\rho}+\frac{1}{2}\mathbf{v}\cdot\mathbf{v},
\end{equation}
with $\gamma$ being the adiabatic index, calculated as the ratio of the constant-pressure heat capacity to the constant-volume heat capacity of the gas.

Let $\mathbf{q}=(\rho,v^1,v^2,v^3,p)$ denotes the hydrodynamic primitive variable, $\mathbf{Q}=(\rho,\rho v_1, \rho v_2, \rho v_3,\rho e)$ the conserved variable and $\calbf{F}$, $\calbf{G}$ and $\calbf{H}$ be the flux densities in three dimensions respectively. It is illustrative to write their components explicitly as:
\begin{eqnarray}
    \calbf{F} & = & (\rho v^1, \rho v^1v_1 + p, \rho v^1v_2, \rho v^1v_3, \rho v^1 h) \label{eqn:rho1} \\
    \calbf{G} & = & (\rho v^2, \rho v^2v_1, \rho v^2v_2 + p, \rho v^2v_3, \rho v^2 h) \\
    \calbf{H} & = & (\rho v^3, \rho v^3v_1, \rho v^3v_2, \rho v^3v_3 + p, \rho v^3 h) \label{eqn:rhoe1},
\end{eqnarray}
whereis $h$ is the sum of the specific enthalpy and kinetic energy, and for an ideal gas
\begin{equation}
    h = e+ \frac{p}{\rho} = \frac{\gamma}{\gamma-1}\frac{p}{\rho}+\frac{1}{2}v_\beta v^\beta,
\end{equation}
Substituting the fluxes in equations (\ref{eq:mom}), (\ref{eqn:rho}), (\ref{eqn:rhoe}) by equations (\ref{eqn:rho1}) - (\ref{eqn:rhoe1}), we arrive at a compact form of the Euler equations in the curvilinear coordinate system:
\begin{equation}
    \frac{\partial (\sqrt{g}\mathbf{Q})}{\partial t} 
        + \frac{1}{Z_1}\frac{\partial(\sqrt{g}\calbf{F})}{\partial X^1} 
        + \frac{1}{Z_2}\frac{\partial(\sqrt{g}\calbf{G})}{\partial X^2}
        + \frac{1}{Z_3}\frac{\partial(\sqrt{g}\calbf{H})}{\partial X^3} = \sqrt{g}\bm{\Phi}(\mathbf{q}),
        \label{eqn:fv}
\end{equation}
where $\bm{\Phi}$ represents various forcing terms on the right-hand-side of equations (\ref{eq:mom}), (\ref{eqn:rho}), (\ref{eqn:rhoe}).

\subsection{Cubed-sphere finite volume discretization}

In the finite volume method, we definite integrals of a finite volume ($V_{ijk}$)
and three face areas ($A_{i-1/2,j,k}, B_{i,j-1/2,k}, C_{i,j,k-1/2}$) as:
\begin{eqnarray}
    \Delta V_{ijk} &=& \iiint_{V_{ijk}}\sqrt{g}Z_1Z_2Z_3dX^1dX^2dX^3 \label{eqn:V} \\
    \Delta A_{i-1/2,j,k} &=& \iint_{A_{i-1/2,j,k}}\sqrt{g}Z_2Z_3dX^2dX^3 \label{eqn:areaA1} \\
    \Delta B_{i,j-1/2,k} &=& \iint_{B_{i,j-1/2,k}}Z_1Z_3dX^1dX^3 \label{eqn:areaB1} \\
    \Delta C_{i,j,k-1/2} &=& \iint_{C_{i,j,k-1/2}}Z_1Z_2dX^1dX^2. \label{eqn:areC1}
\end{eqnarray}
There is a $\sqrt{g}$ factor in equations (\ref{eqn:V}) and (\ref{eqn:areaA1}) because the $X^2$-axis and $X^3$-axis are at an angle $\Theta$, while the $X^1$ axis and $X^3$ axis are orthogonal to each other. An illustration of the relation of the three directions can be found in Figure \ref{fig:geometry}.

Similarly, we can define the volume-weighted average of the conserved variable as:
\begin{equation}
    \mathbf{Q}_{ijk} = \frac{1}{\Delta V_{ijk}}\iiint_{V_{ijk}}
    \mathbf{Q}\sqrt{g}Z_1Z_2Z_3dX^1dX^2dX^3,
\end{equation}
and the area-weighted average of the normal fluxes:
\begin{eqnarray}
    \calbf{F}^\bot_{i-1/2,j,k} &=& \frac{1}{\Delta A_{i-1/2,j,k}}\iint_{A_{i-1/2,j,k}}
    \calbf{F}\sqrt{g}Z_2Z_3dX^2dX^3 \\
    \calbf{G}^\bot_{i,j-1/2,k} &=& \frac{1}{\Delta B_{i,j-1/2,k}}\iint_{B_{i,j-1/2,k}}
    \calbf{G}\sqrt{g}Z_1Z_3dX^1dX^3 \\
    \calbf{H}^\bot_{i,j,k-1/2} &=& \frac{1}{\Delta C_{i,j,k-1/2}}\iint_{C_{i,j,k-1/2}}
    \calbf{H}\sqrt{g}Z_1Z_2dX^1dX^2.
\end{eqnarray}
We would like to emphasize that the fluxes:
\begin{equation}
    \calbf{F}^\bot_{i-1/2,j,k},\; \calbf{G}^\bot_{i,j-1/2,k}, \; \calbf{H}^\bot_{i,j,k-1/2},
\end{equation}
are perpendicular to each face area of the finite volume. This allows the use of any Riemann Solver that is supplied by the base model without rewriting a new one for the non-orthogonal geometry.

Finally, we use the superscripts $\{n,n+1,*\}$ to denote discrete time steps. The finite volume discretization of equation (\ref{eqn:fv}) with explicit spatial time stepping reads as:
\begin{equation}
\begin{aligned}
    \frac{\mathbf{Q}^{n+1}_{ijk}-\mathbf{Q}^{n}_{ijk}}{\Delta t}
    & + \frac{\calbf{F}^{\bot,*}_{i+1/2,j,k}\Delta A_{i+1/2,j,k}
        -\calbf{F}^{\bot,*}_{i-1/2,j,k}\Delta A_{i-1/2,j,k}}{\Delta V_{ijk}} \\
    & + \frac{\calbf{G}^{\bot,n}_{i,j+1/2,k}\Delta B_{i,j+1/2,k}
        -\calbf{G}^{\bot,n}_{i,j-1/2,k}\Delta B_{i,j-1/2,k}}{\Delta V_{ijk}} \\
    & + \frac{\calbf{H}^{\bot,n}_{i,j,k+1/2}\Delta C_{i,j,k+1/2}
        -\calbf{H}^{\bot,n}_{i,j,k-1/2}\Delta C_{i,j,k-1/2}}{\Delta V_{ijk}} = \bm{\Phi}^*_{ijk}, \label{eqn:final}
\end{aligned}
\end{equation}
where the normal flux densities $\calbf{F}^\bot$, $\calbf{G}^\bot$, $\calbf{H}^\bot$ are evaluated at some time between $n$ and $n+1$ depending on the time integration scheme. It is clear at this moment that the finite volume model in a cubed-sphere geometry bears the same numerical scheme as any Cartesian model. This feature makes the transformation of a Cartesian model to a spherical model extremely easy if one correctly calculates the normal fluxes using a Riemann Solver and the finite volume and areas have their usual geometry meaning (facilitated by unit bases vectors). 

The flux density can also be evaluated implicitly using a prediction-correction method, in which a prediction flux density is calculated using any explicit method, and the implicit correction flux density is obtained by linearization. Implicit flux densities are usually employed for the vertical dimension of a GCM because the vertical grid spacing is much smaller than the horizontal grid spacing. In ExoCubed, we use the VIC scheme developed by \cite{ge2020global} to solve the implicit flux density for the vertical dimension only, denoted as $\calbf{F}^{\bot,*}$ and use the explicit flux densities for two horizontal directions. Because the vertical direction is perpendicular to the two horizontal directions, no change is needed for the implicit scheme.

\subsection{Projection and de-projection in Riemann Solvers}
Given the left/right reconstructed primitive variables ($\mathbf{q}^-,\mathbf{q}^+$), the Riemann problem is usually solved in an orthonormal basis frame, which yields the fluxes normal to the face boundary between two finite volume cells. Since the coordinate line is not orthogonal to the face boundary in the horizontal directions in the cubed-sphere geometry, we project the primitive variables to a local orthonormal basis which is defined by the direction of the face boundary and the vertical direction. The projection is a specialization of solving the general relativistic MHD problem \citep{white2016extension} but generalizes a two-dimensional projection \citep{ullrich2010high} designed for the shallow water model to three dimensions.

Denote three basis vectors of the orthonormal basis frame at the cell interface as $\{\mathbf{e}_1,\mathbf{e}_2,\mathbf{e}_3\}$. The first basis vector $\mathbf{e}_1$ is the same as the one in the original basis, $\mathbf{b}_1$:
\begin{equation}
    \mathbf{e}_1=\mathbf{b}_1,
    \label{eq:e1b1}
\end{equation}
since $\mathbf{b}_1$ is already normalized, we do not need further normalization. If the face boundary is spanned by $\{\mathbf{b}_1,\mathbf{b}_3\}$ ($X^2$-face), then its normal direction is the same as the contra-variant basis, $\mathbf{b}^2$ because:
\begin{equation}
    \mathbf{b}^2\cdot\mathbf{b}_1 = \mathbf{b}^2\cdot\mathbf{b}_3=0.
\end{equation}
So, the second basis vector is defined as the normalized $\mathbf{b}^2$ such that:
\begin{equation}
    \mathbf{e}_2=\mathbf{b}^2/|\mathbf{b}^2|.
\end{equation}
The third basis vector is the same as $\mathbf{b}_3$ because $\mathbf{b}_3$ is normalized:
\begin{equation}
    \mathbf{e}_3=\mathbf{b}_3.
\end{equation}
The orthonormal basis frame for the $X^3$-face can be constructed similarly using $\mathbf{b}^3$ as the direction for the face normal.

Let $\mathbf{T}$ be the velocity or flux density vector. Component-wise relations between the original basis and the orthonormal basis are:
\begin{equation}
    \mathbf{T}=T^i\mathbf{b}_i=(\text{O}_2\mathbf{T})^i\mathbf{e}_i.
\end{equation}
The components of the transformed vector $\text{O}_2\mathbf{T}$ are obtained by the inner product of the $\mathbf{T}$ vector and the new basis vectors:
\begin{eqnarray}
    (\text{O}_2\mathbf{T})^1 &=& \mathbf{T}\cdot\mathbf{e}^1=T^1 \\
    (\text{O}_2\mathbf{T})^2 &=& \mathbf{T}\cdot\mathbf{e}^2=T^2/\sqrt{g^{22}} \\
    (\text{O}_2\mathbf{T})^3 &=& \mathbf{T}\cdot\mathbf{e}^3=(T^2g_{23}+T^3g_{33})/\sqrt{g_{33}}
\end{eqnarray}
So the projection operator $\text{O}_2$ can be written in the matrix form as:
\begin{equation}
\text{O}_2=
\begin{bmatrix}
1 & 0 & 0 \\
0 & 1/\sqrt{g^{22}} & 0\\
0 & g_{23}/\sqrt{g_{33}} & \sqrt{g_{33}}
\end{bmatrix},
\label{eqn:O2}
\end{equation}
and its inverse is
\begin{equation}
\text{O}_2^{-1}=
\begin{bmatrix}
1 & 0 & 0 \\
0 & \sqrt{g^{22}} & 0\\
0 & -\sqrt{g^{22}}g_{23}/g_{33} & 1/\sqrt{g_{33}}
\end{bmatrix}.
\end{equation}
The component-wise de-projection formula reads as:
\begin{eqnarray}
    (\text{O}_2^{-1}\mathbf{T})^2 &=& T^2\sqrt{g_{22}} \\
    (\text{O}_2^{-1}\mathbf{T})^3 &=& -T^2\sqrt{g^{22}}g_{23}/g_{33}+T^3/\sqrt{g_{33}}
\end{eqnarray}
Similarly, the projection O$_3$ is
\begin{equation}
\text{O}_3=
\begin{bmatrix}
1 & 0 & 0 \\
0 & \sqrt{g_{22}} & g_{32}/\sqrt{g_{22}}\\
0 & 0 & 1/\sqrt{g^{33}}
\end{bmatrix},
\label{eqn:O3}
\end{equation}
and its inverse:
\begin{equation}
\text{O}_3^{-1}=
\begin{bmatrix}
1 & 0 & 0 \\
0 & 1/\sqrt{g_{22}} & -\sqrt{g^{33}}g_{32}/g_{22}\\
0 & 0 & \sqrt{g^{33}}
\end{bmatrix},
\label{eq:O3-1}
\end{equation}
Note that the vertical component is unaffected by the projection since the vertical axis in the orthonormal frame is the same as that in the original frame. The projection matrix $\text{O}_2$ is also derived by \cite{ullrich2010high} in a two-dimensional shallow water manner.

\begin{figure}
    \centering
    \includegraphics[width=0.7\textwidth]{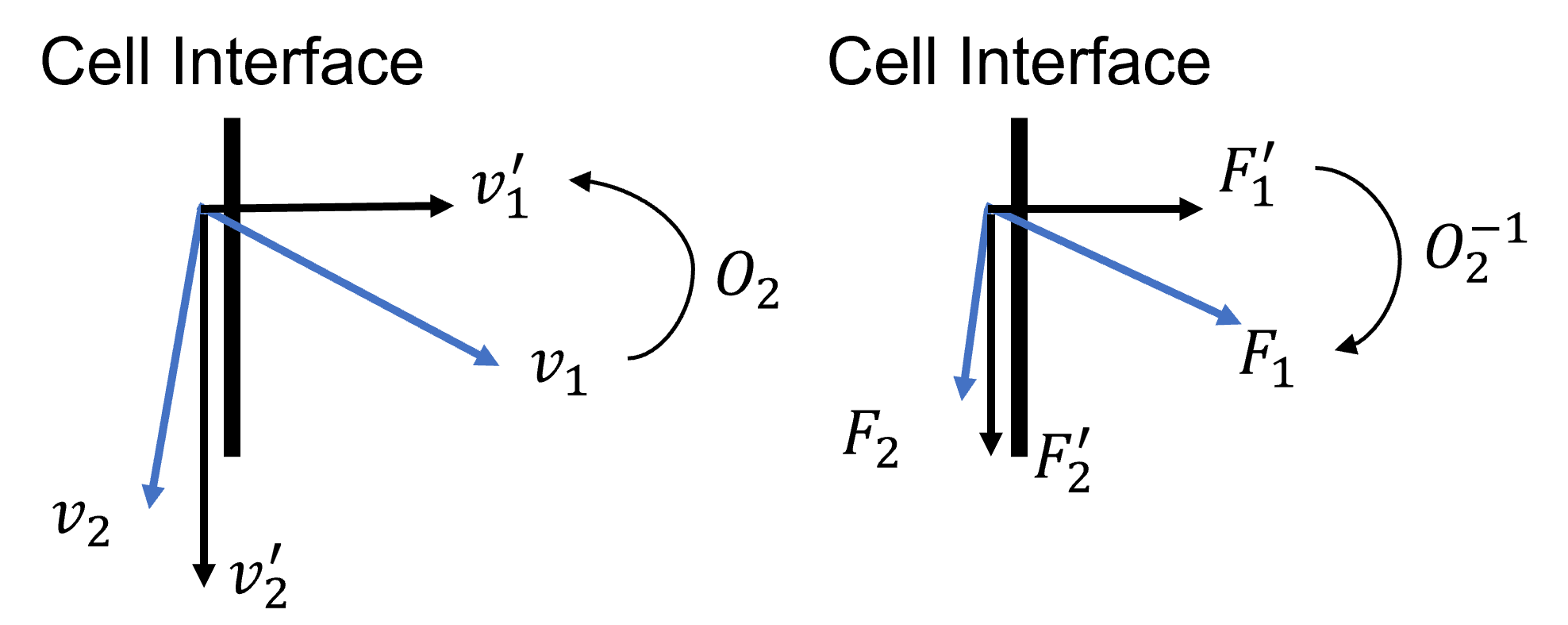}
    \caption{Illustration of the steps of the projection operations for Riemann Solver calculations. We convert the velocity on the covariant basis to the local orthonormal basis before the Riemann Solver calculation (left) and convert the flux from the local orthonormal basis to the covariant basis.}
    \label{fig:rsolver}
\end{figure}

Now we describe the whole process of projection, Riemann Solver, and de-projection using $X^2$-face as an example. A corresponding illustration of this process can be found in Figure \ref{fig:rsolver}. First, we obtain the left and right components of primitive variables in the original basis at the face boundary by a reconstruction method based on cell-averaged values, denoted as:
\begin{equation}
    \text{Step 1:}\quad \mathbf{q}^{\pm}_{j-1/2}=(\rho^\pm,v^{1,\pm},v^{2,\pm},v^{3,\pm},p^\pm)_{j-1/2}.
    \label{eq:riemann1}
\end{equation}
Here $j-1/2$ refers to the left face of a finite volume (See Fig. \ref{fig:cubed-grid}).
Second, the projected components in the local orthonormal basis are obtained by the projection matrix (\ref{eqn:O2}):
\begin{equation}
    \text{Step 2:}\quad \text{O}_2\mathbf{q}^{\pm}_{j-1/2}=\text{O}_2\cdot \mathbf{q}^{\pm}_{j-1/2},
\end{equation}
where ``$\cdot$" indicates a matrix multiplication. Step 2 yields two primitive variables, $\text{O}_2\mathbf{q}^{\pm}_{j-1/2}$, that are defined in the local orthogonal reference frame.
Third, we solve the orthonormal Riemann problem in the local orthonormal frame such that the resulting flux densities are:
\begin{equation}
    \text{Step 3:}\quad \text{O}_2\calbf{G}_{j-1/2}^\bot=\text{RiemmanSolver}(\text{O}_2\mathbf{q}^-_{j-1/2},\text{O}_2\mathbf{q}^+_{j-1/2}).
\end{equation}
Step 3 yields the normal flux densities across the cell interface face but the components are defined in the local orthogonal reference frame.
Finally, we de-project the normal flux density from the local orthonormal frame to the original reference frame:
\begin{equation}
    \text{Step 4:}\quad \calbf{G}^\bot_{j-1/2}
    = \text{O}_2^{-1}\text{O}_2\calbf{G}^\bot_{j-1/2}
    = \text{O}_2^{-1} \cdot \text{O}_2\calbf{G}^{\pm}_{j-1/2}.
    \label{eq:riemann4}
\end{equation}
Step 4 yields normal flux densities that are expressed in the cubed-sphere reference frame. The result of step 4 is then supplied to equation ($\ref{eqn:fv}$) to calculate the flux divergence.

We have verified the validity of the projection and de-projection algorithm using a simple circular breaking dam simulation using the affine coordinates, benchmarked by the simulation results in Cartesian coordinates (not shown).

\subsection{Velocity rotation and interpolation}
\begin{figure}
    \centering
    \includegraphics[width=0.5\textwidth]{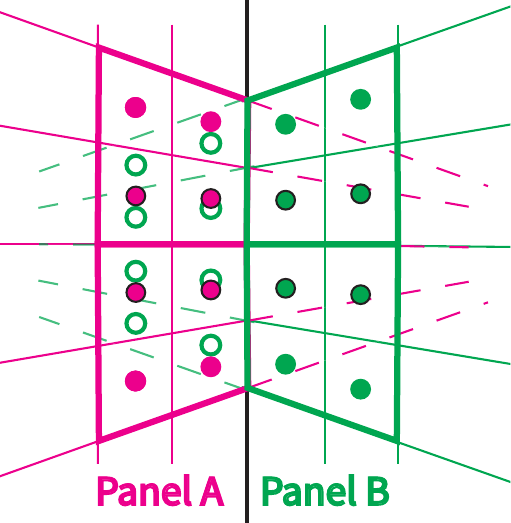}
    \caption{Grid visualization of ghost cells at the panel boundary. Panel A's coordinate lines and cells are depicted in pink, while Panel B's are in green. The delineating boundary between Panels A and B is shown in black. Cell centers are represented by dots. Ghost cells in Panel B, designated by open green circles, will be populated using cell-averaged values from Panel A. Bold pink and green lines indicate meshblock boundaries. Within each panel, four dots bordered in black can be observed: the lower two dots correspond to the top meshblock's intra-panel ghost zones and the upper two to the bottom meshblock's. Values in the open green circles are determined through interpolation from the pink dots column by column from left to right.}
    \label{fig:in-panel-sync}
\end{figure}

Crossing the panel boundaries, the ghost cells of one panel do not align with the interior cells of the neighboring panel. So, when the values in the interior cells of the neighboring panel are used to fill the values in the ghost cells of the current panel, interpolation is necessary (Figure \ref{fig:in-panel-sync}). In the gnomonic equiangular cubed-sphere geometry, this problem is simplified to a one-dimensional linear interpolation problem because each coordinate line is on a great circle of the sphere. Denoting the positions as angular positions along the arcs. The angular position is calculated as
\begin{equation}
    \alpha_n = \cos^{-1}\left(\sqrt{\frac{1+\tan^2\left(\frac{N-1-2k}{4N}\pi\right)}{\tan^2\left(\frac{N-1-2k}{4N}\pi\right)+\tan^2\left(\frac{2n+1}{4N}\pi\right)}}\right)
    \label{eqn:interp}
\end{equation}
Here $N$ is the number of cells along the direction of interpolation, and $n$ is the corresponding index of the cell. $k$ denotes the distance of the cells from the boundaries. When $k<0$ or $k\geq N$, the cell is within the ghost zone.
\newline
The interpolation happens in a matching manner. To interpolate the values on ghost zones $(k<0)$, the locations for interior points $k'=-k-1$ are used. Note that Equation (\ref{eqn:interp}) is universal and has the same form for all the panel boundaries.

Not only do the values in the ghost cells need to be interpolated, the velocity components also need to be transformed since the coordinate lines are not continuous crossing panel boundaries.

To transform a vector from $\{\mathbf{b}_1,\mathbf{b}_2,\mathbf{b}_3\}$ (basis in the source panel) to $\{\mathbf{b}_1,\mathbf{b}_2',\mathbf{b}_3'\}$ (basis in the target panel), we calculate the projection matrix as
\begin{equation}
\text{O}=R^n
\begin{bmatrix}
1 & 0 & 0\\
0 & 1 & \cos\Theta-\cot\Theta' \sin\Theta\\
0 & 0 & \sin\Theta/\sin\Theta'
\end{bmatrix}
\label{eqn:decomp_mat}
\end{equation}
Here $R$ is the rotation matrix that rotates the local coordinate system counterclockwise, and
\begin{equation}
\text{R}=
\begin{bmatrix}
1 & 0 & 0\\
0 & 0 & -1\\
0 & 1 & 0
\end{bmatrix}
\end{equation}
The number of such rotations needed, i.e. the value of $n$ in each panel boundary is summarized in the following table:

\begin{table}[H]
    \centering
    \begin{tabular}{|c||*{6}{c|}}\hline
    
    \backslashbox{Target direction}{Source panel}
    &\makebox[3em]{1}&\makebox[3em]{2}&\makebox[3em]{3} &\makebox[3em]{4}
    &\makebox[3em]{5}&\makebox[3em]{6}\\\hline\hline
    Up & 2 & 0 & 3 & 1 & 0 & 2\\\hline
    Down & 0 & 0 & 1 & 3 & 2 & 2\\\hline
    Left & 1 & 0 & 0 & 0 & 3 & 0\\\hline
    Right & 3 & 0 & 0 & 0 & 1 & 0\\\hline
    \end{tabular}
    \caption{Summary of numbers of rotations needed for transformation on panel boundaries.}
    \label{tab:rot_relation}
\end{table}

For calculation, we need to first find the coordinates of interpolated points in both panels. For the target panel (where the point is within the ghost zone), the grid is apparent, while in the source panel, we can calculate the $y$-coordinate values from the angular position $\alpha_n$:
\begin{equation}
    y=\pm \sqrt{1+(x)^2}\tan \alpha_n
\end{equation}
and
\begin{equation}
    \tan \alpha_n = \big(\frac{1}{\cos^2 \alpha_n}-1\big)^\frac{1}{2}=\big(\frac{\tan^2\left(\frac{2n+1}{4N}\pi\right)-1}{\tan^2\left(\frac{N-1-2k}{4N}\pi\right)+1}\big)^\frac{1}{2}
\end{equation}
The sign of the $y$-coordinate abides with the sign in the target panel. Note that we have assumed the panel boundary is along $y$-direction in this case, but for boundaries along any direction the interpolation is similarly treated. After this step, we use Equations (\ref{eqn:b1}) - (\ref{eqn:b3}) to calculate the local basis vectors and calculate the transformation matrices.

\subsection{Left/right states synchronization}
Since interpolation and rotation are performed to fill the ghost zones outside the panel boundary, the restricted states at the panel interface may be slightly different for the two panels sharing one cell interface. If the left/right states are different, the result of the Riemann Solver will also be different, leading to inconsistent fluxes across the panel boundary. The GFDL's FV3 model solves this issue by averaging the fluxes computed by the left and right panels \citep{chen2021lmars,mouallem2023implementation}. We take a different approach. We implemented a communication function that exchanges the reconstructed left/right states in both panels rather than fluxes. Specifically, the right state at the cell interface gets the right values reconstructed by the right panel, and the left state at the cell interface gets the left values reconstructed by the left panel. This approach minimizes the use of the values in the ghost cell, which are inaccurate. After the exchange is done, the two states at the left/right side of the panel interface are synchronized for both panels. This ensures that the Riemman Solvers in each panel yield the same flux, and thus improves the conservation properties.

\subsection{Metric terms}
Due to the change of the basis vectors, the momentum equation (\ref{eqn:rhov_t}) contains extra metric terms acting as the source term to the local momentum. The expression of the metric terms depends on the type of coordinate system, the choice of basis vectors and the variables to solve. In the classic geophysical fluid dynamics textbook \citep{pedlosky2013geophysical}, the metric terms only involve the components of the Reynolds stress tensor because the differential equations are written in primitive variables. Since we solve the conservative form of the equations, our metric terms are calculated from the total stress tensor which includes the contribution from the pressure. In spherical-polar coordinates, the metric terms are:
\begin{eqnarray}
    M^1 & = & \frac{2p}{r} + \frac{\rho(v^2v^2+v^3v^3)}{r} \label{eqn:M1a} \\
    M^2 & = & -\frac{\rho v^1v^2}{r} + \frac{(p+\rho v^3v^3)\cot\theta}{r} \label{eqn:M2a} \\
    M^3 & = & -\frac{\rho v^1v^3}{r} - \frac{\rho v^2v^3\cot\theta}{r}, \label{eqn:M3a}
\end{eqnarray}
where $\theta$ is the polar angle and $v^1$, $v^2$, $v^3$ are velocities in the radial, polar and azimuthal directions respectively. In cubed-sphere geometry, the metric terms are:

\begin{equation}
    M_\alpha = \Gamma^\gamma_{\alpha\beta}T^\beta_\gamma \label{eqn:ma}
\end{equation}

Substitute equations (\ref{eqn:gamma1}) - (\ref{eqn:gamma2}) and (\ref{eqn:tab}) into equation (\ref{eqn:ma}), we derive the metric terms in the cubed-sphere geometry:
\begin{eqnarray}
    M_1 & = & \frac{2p}{r} + \frac{\rho(v^2v_2+v^3v_3)}{r} \label{eqn:M1b} \\
    M_2 & = & - \frac{\rho v^1 v_2}{r}
    -\frac{x}{rD}\big(p + \rho v^3 v^3 (\sin\Theta)^2\big) 
    \label{eqn:M2b} \\
    M_3 & = & - \frac{\rho v^1 v_3}{r}
    -\frac{y}{rC}\big(p + \rho v^2 v^2 (\sin\Theta)^2\big). \label{eqn:M3b}
\end{eqnarray}

Equations (\ref{eqn:M1b}) - (\ref{eqn:M3b}) can be compared directly to equations (\ref{eqn:M1a}) - (\ref{eqn:M3a}). First, we find that a kinetic energy term ($v^2v_2 + v^3v_3$) appear in both equations (\ref{eqn:M1a}) and (\ref{eqn:M1b}). Second, the factor, $\cot\theta$, represents the change of face area in the spherical polar geometry whereas, the factors $x/(rD)$ and $y/(rC)$ represent the change of face area in the cubed-sphere geometry. Metric terms in horizontal directions are more symmetric in the cubed-sphere geometry than those in the spherical-polar geometry due to the symmetric nature of the coordinate system.

\section{Benchmark Tests}
In this section, we present tests to evaluate the performance of the dynamic core. From the benchmark simulations, we can draw conclusions on the accuracy, efficiency, and stability of the cubed-sphere model presented in this work. We also evaluate how well the different physical quantities are conserved. 

Tests range from shallow water tests (2D global models) to 3D general circulation models. In 3D general circulation model tests, we applied the Vertical Implicit Correction (VIC) scheme \citep{ge2020global} to further improve the speed of simulations. It is worth noting that, with VIC applied, the relationship between the time steps and Courant–Friedrichs–Lewy (CFL) number is calculated as $\Delta t_{VIC}=\text{CFL}\times\min (\Delta x, \Delta y)/c_s$ since the implicit scheme is applied along the $X^1$-axis and is unconditionally stable.

\subsection{Hydrodynamic blast in one cubed-sphere panel}
\begin{figure}
    \centering
    \includegraphics[width=0.8\textwidth]{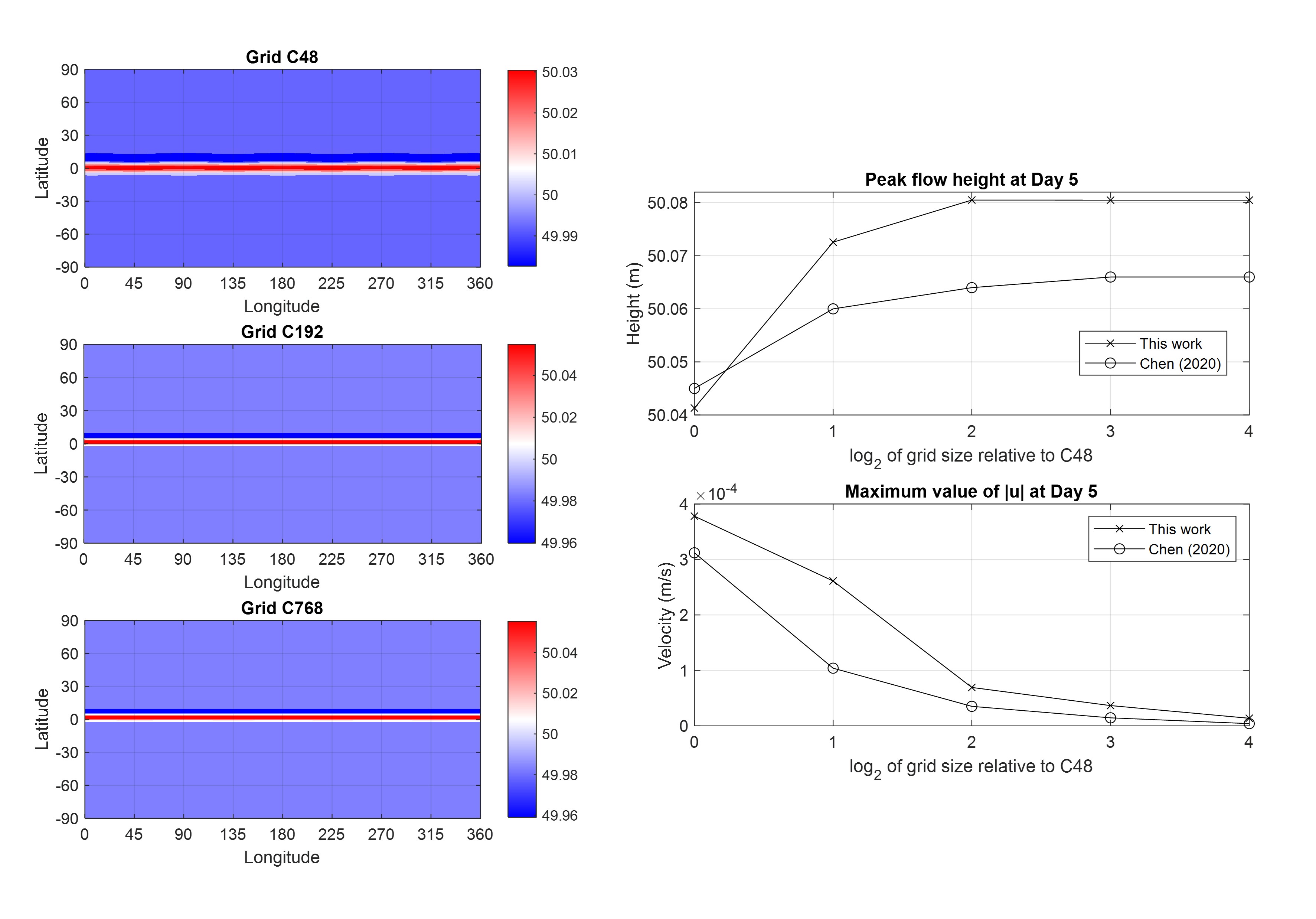}
    \caption{The flow height field at day 5 (left), and the convergence of the peak flow height and maximum zonal flow speed at day 5 when we increase our grid resolution (right).}
    \label{fig:blast}
\end{figure}

For the first test, the experiment used in \citet{chen2021lmars} is adopted to validate the dispersion and dissipation properties. This test involves simulating a splash of a sinusoidal droplet on a non-rotational sphere. The initial perturbation to a uniform height field is limited to a small region at the pole. The initial geopotential height is written as:
\begin{equation}
\Phi_{init}=
\left\{
\begin{aligned}
& \Phi_0 + \Phi'\cos{\frac{\pi r}{2R}}, \text{ if }r<R \\
& \Phi_0, \text{ otherwise}
\end{aligned}
\right.
\end{equation}
Here $r$ is the distance from the north pole, $R$ is 500 km, $\Phi_0=\text{50 m}\times g$, and $\Phi'=\text{1 m}\times g$. These values give a gravity wave that reaches the south pole in 10 days. In Figure \ref{fig:blast}, the height fields at day 5, when the wave reaches the equator, are shown in the left column.

For the tests, we conducted the simulation in multiple different horizontal resolutions. In this paper, we use the letter ``C" followed by a number, which is the number of grids along each direction for each of the six panels, to label the resolution of each configuration. For example, ``C48" denotes the resolution of 48$\times$48$\times$6 on the surface of the sphere.

The quantitative comparison between grids and \citet{chen2021lmars} (gnomonic equiangular coordinate with rk3 scheme results are shown here) is shown in the right column of Figure \ref{fig:blast}. As expected, the solutions are less dissipative and preserve the spherical symmetry better by having a smaller zonal wind magnitude with finer grids. Intriguingly, our cubed-sphere model is less dissipative in comparison to \citet{chen2021lmars} at finer grids.

\subsection{Rossby-Haurwitz Wave and Steady Zonal Jets}
\citet{williamson1992standard} provided a widely-used standard test set for testing the accuracy of shallow water equations on spherical geometry. For shallow-water tests, we performed tests on test cases 2 (global steady state nonlinear zonal geostrophic flow) and 6 (Rossby-Haurwitz Wave) to evaluate the order of accuracy of our model. Note that in \citet{williamson1992standard}, there are four different configurations for case 2 with jets going along different directions, but here we show only one case, where the jet is tilted 0.05 radians from the perfectly zonal direction, since the results from all four configurations are essentially similar.

\begin{figure}
    \centering
    \includegraphics[width=1.0\textwidth]{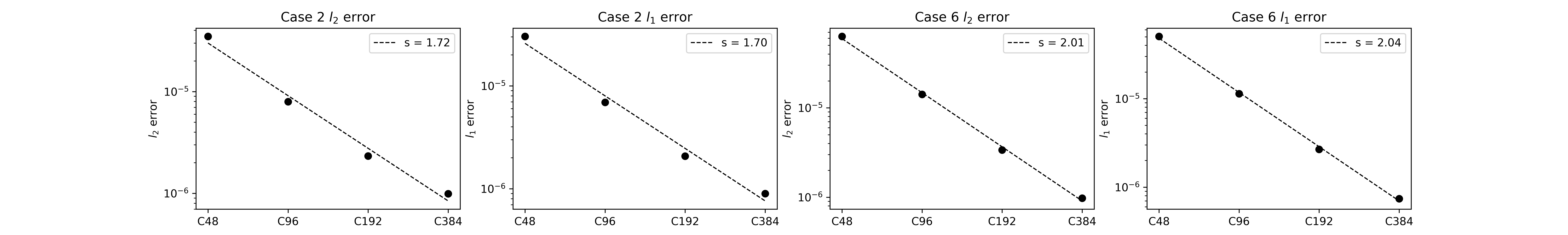}
    \caption{The average $l_1$ and $l_2$ error relative to high-resolution solution for cases 2 and 6 \citep{williamson1992standard}. The log-linear fits of the error values are plotted, and the slopes are labeled.}
    \label{fig:W92_error}
\end{figure}

These two tests are conducted with different resolutions, from C48 to C384, and we use the results from C768 to calculate the normalized global errors $l_1$ and $l_2$ in layer thickness:
\begin{equation}
    l_1(h) = \frac{I\left[ |h\left(\lambda,\theta\right)-h_T\left(\lambda,\theta\right)|\right]}{I\left[|h_T\left(\lambda,\theta\right)|\right]}
\end{equation}
\begin{equation}
    l_1(h) = \frac{\left\{I\left[ \left(h\left(\lambda,\theta\right)-h_T\left(\lambda,\theta\right)\right)^2\right]\right\}^{\frac{1}{2}}}{\left\{I\left[\left(h_T\left(\lambda,\theta\right)\right)^2\right]\right\}^{\frac{1}{2}}}
\end{equation}
Here $h_T$ is the true solution, and $I[]$ is the operator that integrates over the surface of the sphere. The respective results are shown in Figure \ref{fig:W92_error}. The log-linear fits are shown in the plots, and the slope values correspond to the order of accuracy. In the tests, we used CFL=0.9, while we tested CFL=0.1 and found no significant difference in both errors and the fitted slope values. This insensitivity to a nine-fold difference in time step has important implications for the 3D simulations. The vertical implicit scheme ensures unconditional stability in the vertical direction, albeit at the cost of damping fast waves like sound waves. Yet, for long-period waves like Rossby waves, the time step is significantly smaller than the wave period, resulting in minimal distortion of wave propagation by the vertical implicit scheme. Non-hydrostatic models with vertical implicit schemes should resolve slow waves as well as any hydrostatic models. This contention is validated by the findings in \citet{ge2020global}, where 3D simulations with the implicit scheme turned on and off showed no discernible differences. This supports the use of larger time steps enabled by the vertical implicit scheme to enhance computational efficiency.

We found that the model's convergence order in the two test cases is about 2, even though the 5-th order numerical scheme (WENO5) is applied in the reconstruction. Currently, the convergence order is limited by the second-order cell averaging scheme when converting between the conserved and primitive variables. \citet{stone2020athena++} similarly reported second order of accuracy in general, even if a Piecewise Parabolic Method (PPM) is used for the reconstruction. While the second-order cell averaging scheme poses a second-order convergence rate to the whole model, applying higher-order reconstruction schemes has the advantage of reducing both the amplitude and the phase error at a fixed model resolution. Specifically, in the hydrodynamic blast test, we noted that the lower-order reconstruction schemes are more dissipative than higher-order schemes such as WENO5. Since the viscosity of the atmospheric motion is small, using a higher-order reconstruction scheme is essential for resolving hydrodynamic instability in a weakly dissipative system. For example, in a standard rising bubble test \citep{Robert1993Bubble}, where the temperature anomaly is only 0.5 K,  a 5th-order reconstruction scheme is necessary to resolve the Kelvin-Helmholtz instability caused by the rising bubble. Yet, in the standard sinking bubble test \citep{Straka1993Numerical}, where the temperature anomaly is about 20 K, a 2nd-order reconstruction scheme suffices to resolve the instability. 

\begin{figure}
    \centering
    \includegraphics[width=1.0\textwidth]{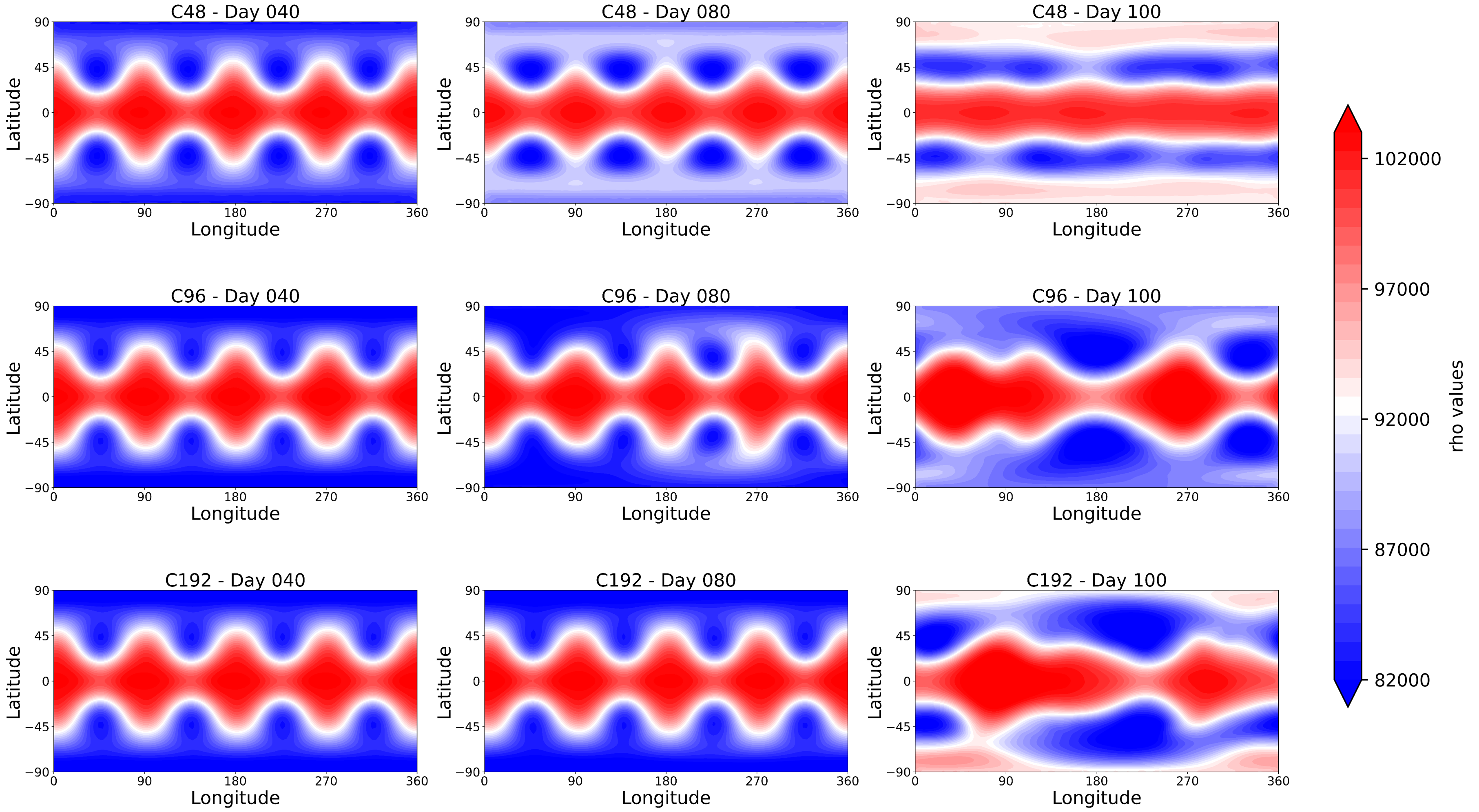}
    \caption{The flow height field at day 40 (left column), day 80 (middle column), and day 100 (right column) for resolutions C48 to C192.}
    \label{fig:W92}
\end{figure}

In Figure \ref{fig:W92}, we show the height field generated by different levels of resolutions, at days 40, 80, and 100, respectively. The Rossby-Haurwitz wave is supposed to be a wave that travels from east to west without changing the shape, but the error accumulates, and the wave breaks up at later dates. For finite volume models, the truncation error makes the wave more susceptible to dynamic instability, and it usually breaks up at about days 30-40 \citep{Smith2006Revisiting, Thuburn2000Numerical}. The spectral semi-Lagrangian models have a chance to keep stable till around day 100 \citep{Thuburn2000Numerical}. In Figure \ref{fig:W92}, even the coarsest grid C48 survives beyond day 80, and for various resolutions, the cases all become unstable at about day 90-100. Such solutions show a slower error accumulation than other finite volume methods previously reported.

\subsection{Held-Suarez test of Earth's climate}
The Held-Suarez test of Earth's climate is designed as a common problem for comparisons between different dynamical cores of GCMs \citep{HS94}. Focused on the global-scale atmospheric features and the long-term state of an earth-like planet, the comparisons are usually made on a final quasi-steady state, and averaged quantities are used. The final quasi-equilibrium state of this test presents a latitudinal temperature gradient and large scale winds. This is similar to the atmosphere of the earth.

In addition to the Coriolis force corresponding to the rotation rate of the earth, the setup of this simulation involves two major forcing mechanisms. A reference thermal structure is imposed, and the atmosphere is relaxed to the reference structure by a Newtonian cooling scheme. This forcing simplifies the radiative forcing processes:
\begin{equation}
    F_{rad}=-k_T(\phi,\sigma)\rho c_v \left[ T-T_eq(\phi,\sigma)\right]
\end{equation}
Following \citet{HS94}, we use the normalized pressure scale $\sigma=p/p_0$ here, and $p_0=1 \text{bar}$ is the surface pressure. In addition, $c_v$ is the isochoric specific heat of air, and $k_T$ (in the unit of day$^{-1}$) and $T_eq$ (in the unit of K) are the temperature damping strength and the reference thermal structure is defined as
\begin{equation}
    k_T(\phi,\sigma)=0.025+0.225\max\left\{0, \frac{\sigma-0.7}{0.3}\right\}\cos^4\phi
\end{equation}
\begin{equation}
    T_{eq}(\phi,\sigma)=\max\left\{200, \left[ 315-60\sin^2\phi-10\log\sigma\cos^2\phi\right]\sigma^\kappa\right\}
\end{equation}
where $\kappa=R/c_p=2/7$. The Rayleigh drag is applied to the lower boundary to simplify the boundary effects from the interface with the ground. It appears as an additional source term in the momentum equation
\begin{equation}
    \textbf{M}_{fric}=-k_v(\sigma)\rho\textbf{u}
\end{equation}
The parameter $k_v$, the Rayleigh drag strength, is calculated as (in the unit of day$^{-1}$)
\begin{equation}
    k_v(\sigma)=\max\left\{0,\frac{\sigma-0.7}{0.3}\right\}
\end{equation}
The simulation domain is set to have a 40$\times$40$\times$6 resolution along the horizontal direction. This has a similar number of cells as a spherical-polar grid with an angular resolution of 2.8$^{\circ}$ along both longitudinal and latitudinal directions. The domain is 25 km in height, with 40 layers, giving a vertical resolution of 625 meters. Different from \citet{ge2020global,Mendonça2016THOR}, we have not applied a sponge layer at the top of the atmosphere to absorb the vertically propagating waves, as the stability of the model is not compromised by these waves. For the timestep, we use a high CFL number of 0.9 for high efficiency and also highlighting the stability of the dynamic core. Similar to \citet{ge2020global} and the original work, the simulation reached statistical equilibrium at around day 200, and the simulation continued until day 1200. The statistics of the flow fields for the 1000 days between day 200 and day 1200 are presented here. The timestep $\Delta t$ for a spherical-polar grid with 2.8$^{\circ}$ angular resolution is about 25 seconds, while the ExoCubed dynamic core gives 425 seconds, providing an acceleration over an order of magnitude. Delays due to additional MPI communications needed for the flux synchronization steps exist but are negligible.

\begin{figure}
    \centering
    \includegraphics[width=1.0\textwidth]{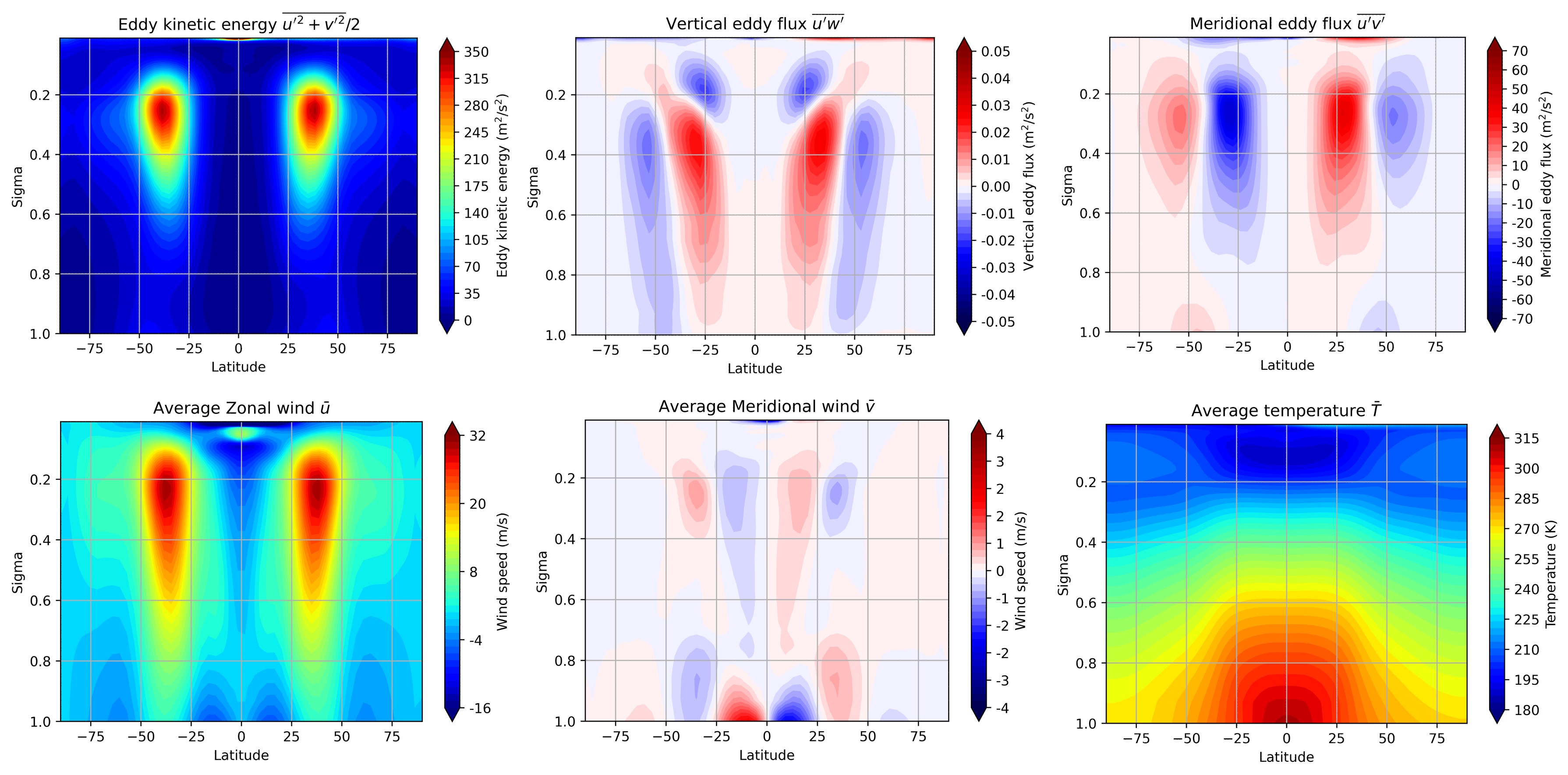}
    \caption{The mean flow fields averaged both zonally and temporally over day 200-1200 for the Held-Suarez test. Note that the $y$-axis is selected to be the normalized pressure scale $\sigma=p/p_0$.}
    \label{fig:hs94}
\end{figure}

The averaged quantities are plotted in Figure \ref{fig:hs94}. From the average meridional wind $\overline{v}$, the Hadley cells and Ferrel cells are clearly defined. The maximum prograde jet has a magnitude of 30.41 m s$^{-1}$, and the maximum meridional wind magnitude is 2.18 m s$^{-1}$. This magnitude of wind maximum is only 1\% from the values reported in \citet{ge2020global}, and the average wind speeds as well as the average temperature structure agree with the results reported in previous studies (e.g. \citet{ge2020global,lin2004vertically,Ullrich2012MCore,Mendonça2016THOR}). In addition to the averaged large-scale circulations, the eddy-related statics are also produced. The meridional and vertical eddy flux combined presents the transport related to the well established Hadley and Ferrel cells. Additionally, the eddy fluxes and the eddy kinetic energy patterns agree with the results in \citet{ge2020global}.

\begin{figure}
    \centering
    \includegraphics[width=0.7\textwidth]{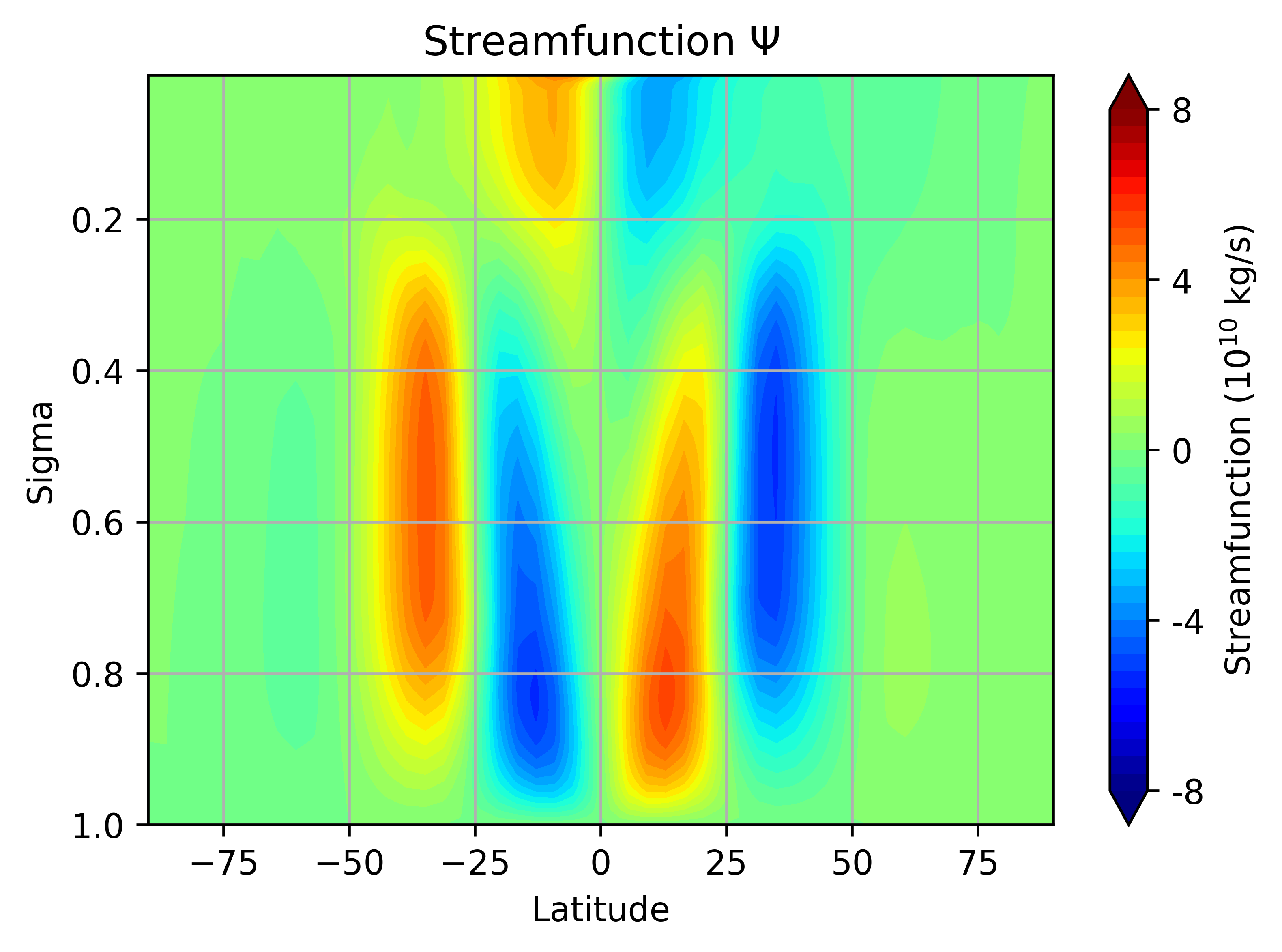}
    \caption{The stream function of Held-Suarez test of the averaged flow field.}
    \label{fig:hs94_psi}
\end{figure}

To better demonstrate the patterns of the general circulation, the streamfunction is calculated based on the averaged velocity fields. There are several pairs of blobs marking major circulation patterns. There are four cells below $\sigma=0.2$, which demonstrate the contours followed by Hadley cells (closer to the equator) and Ferrel cells (farther from the equator). The results agree qualitatively with the results reported by \citet{Mendonça2016THOR}.

A key to note in this model presented is that, at the top of the atmosphere, a sponge layer was not applied, and the top boundary is treated as a solid boundary. This generates an additional pair of convection cells above the Hadley cells, at $\sigma<0.2$. Such structure is seen in both the streamfunction plot (Figure \ref{fig:hs94_psi}) and the meridional eddy flux plot (Figure \ref{fig:hs94}). \citet{Bretherton1969lamb} analytically investigated the scale height of the Lamb wave pattern $H_{Lamb}$, and if we compare it to the pressure scale height $H_p$, the ratio is only associated with the heat capacity ratio $\gamma$:
\begin{equation}
    \frac{H_{Lamb}}{H_p}=\frac{\gamma}{2-\gamma}=2.33
\end{equation}
This agrees with the observation here that the additional cells extend from $\sigma=0.01$ to $\sigma\approx 0.2$, so the results are reasonable interpreted by the lamb waves.

The conservation of the angular momentum, the total energy, and the mass is presented in \ref{fig:stats} (a)-(c). The total AAM about the rotating axis of the planet is calculated as
\begin{equation}
    M=\int_V\left( \Omega R_p \cos \phi + u \right)\rho R_p dV
\end{equation}
Where $R_p$ is the radius of the earth (6371 km). The change of total AAM and the total energy is approximately 1.1\% and 0.2\% respectively over the period of simulation. The change of total mass is 10 orders of magnitude smaller than the initial value. Hence, we conclude that the conservation laws are well followed in this simulation.

\subsection{Shallow Hot Jupiter simulation}
We present a final test with global circulation model on the hot jupiter atmospheres. Observational selection effects lead to most known Jovian-size extrasolar planets being close to their stars. These hot jupiters are likely tidally locked to their host stars, orbiting around them synchronously. This brings about an interesting radiative forcing pattern, as the dayside is always irradiated. This permanent day-night irradiation patterns bring a climate state distinct from those observed in the solar system, making it an interesting problem to simulate. We use this as a test case to validate the capability of our model to simulate an atmosphere that is in a completely different regime. It has been shown that the equator super-rotating jet could have a velocity of about 1000 m s$^{-1}$, which is in the subsonic to sonic regime, and the temperature difference between the day and night sides can exceed 1000 K \citep{Showman2015three}.

\begin{figure}
    \centering
    \includegraphics[width=1.0\textwidth]{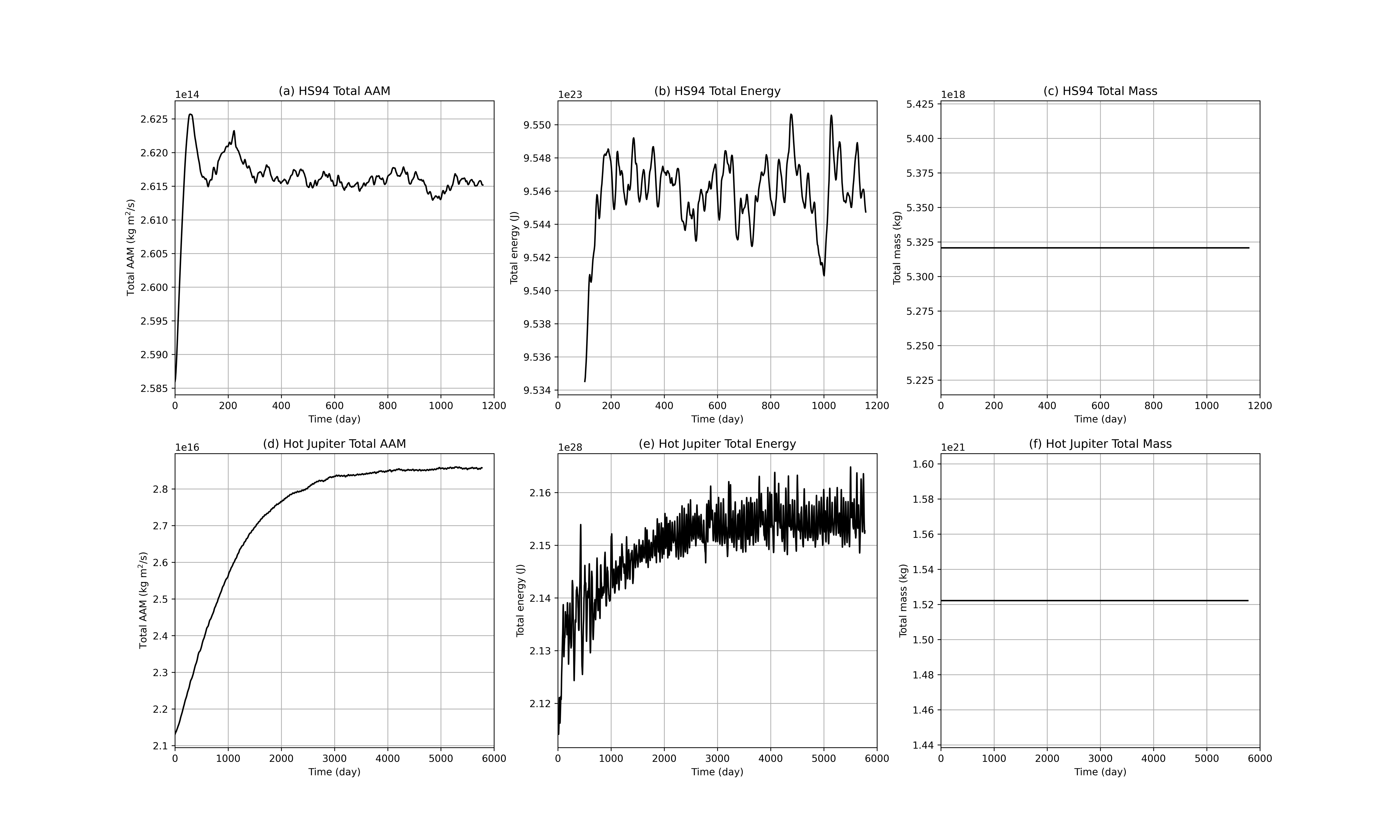}
    \caption{The flow height field at day 5 (left), and the convergence of the peak flow height and maximum zonal flow speed at day 5 when we increase our grid resolution (right).}
    \label{fig:stats}
\end{figure}

The benchmark hot jupiter test is introduced by \citet{Heng2011Atmos}. It follows from a shallow three-dimensional model of the hot jupiter's atmospheric circulation \citep{Menou2009Atmos}. In this simulation, we focus on a planet with radius $R_p=10^5$ km and gravitational acceleration of 8 m s$^{-2}$. The initial bottom pressure is set to be 1 bar, and the initial atmosphere is isothermal at 1600 K. Similar to the Held-Suarez test, the radiative forcing is simulated by relaxing the atmosphere to a prescribed temperature structure with Newtonian cooling, and the reference temperature profile is given by \citep{Menou2009Atmos}
\begin{equation}
    T_eq = T_{vert}+\beta_{trop}\Delta T_{E-P}\cos \lambda \cos \phi
\end{equation}
Here $T_{vert}$ and $\beta_{trop}$ are parameters in the same form as introduced in \citet{Menou2009Atmos}. $T_{E-P}=300$ K is the temperature difference between the equator and the pole. The number of grids is the same as used in the Held-Suarez test, with 40$\times$40$\times$6 in the horizontal direction, and vertically 4500 km with 40 layers. The simulation takes longer (3000 earth days) to reach statistical equilibrium, which is apparent from Figure \ref{fig:stats}. Hence, the average flow fields shown are average values from days 3000-4000.

\begin{figure}
    \centering
    \includegraphics[width=1.0\textwidth]{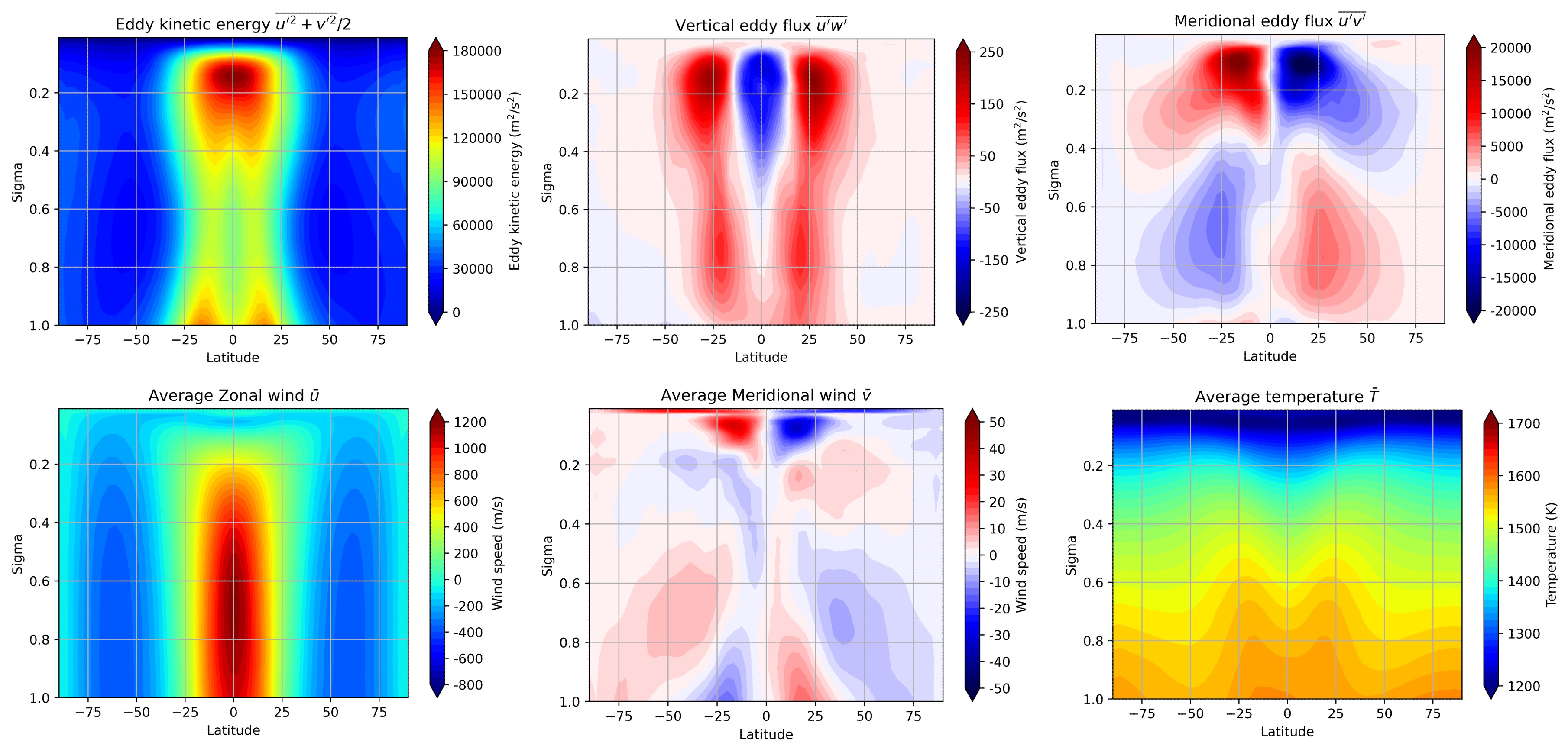}
    \caption{The mean flow fields averaged both zonally and temporally over day 3000-4000 for the hot jupiter test. Note that the $y$-axis is selected to be the normalized pressure scale $\sigma=p/p_0$.}
    \label{fig:hotjupiter}
\end{figure}

Similar to Figure \ref{fig:hs94}, the same set of flow field quantities are shown in Figure \ref{fig:hotjupiter}. The maximum speed of the prograde zonal wind is 1175.13 m s$^{-1}$. The general zonal mean structures, especially the well-reported average temperature and zonal wind profiles, agree well with previous results \citep{ge2020global, Mendonça2016THOR, Heng2011Atmos}. Subtle differences exist, and differences can be attributed to the different capabilities of resolving structures by different grids. 

\begin{figure}
    \centering
    \includegraphics[width=0.7\textwidth]{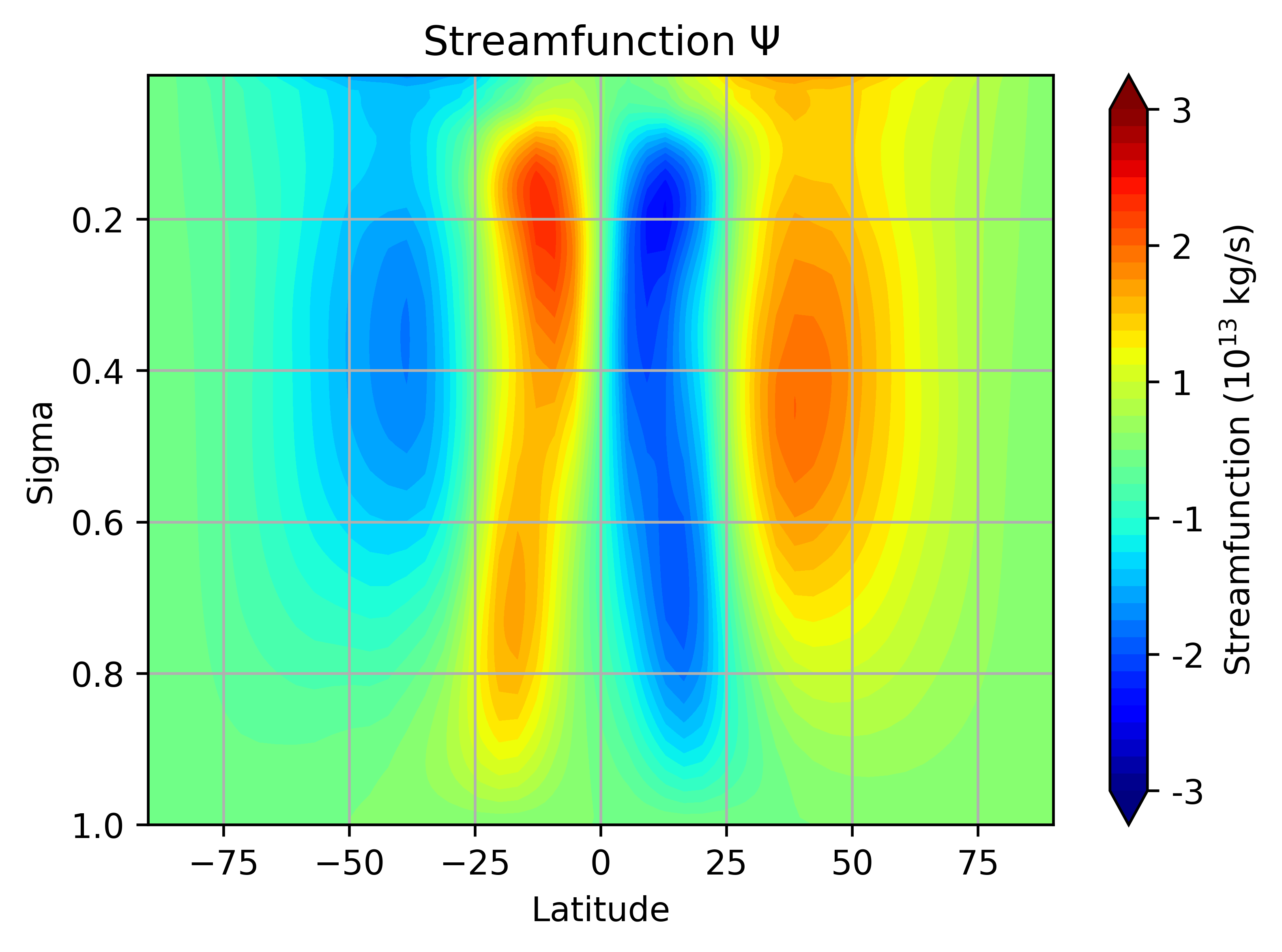}
    \caption{The stream function of shallow hot Jupiter test of the averaged flow field.}
    \label{fig:hotjupiter_psi}
\end{figure}

In Figure \ref{fig:hotjupiter_psi}, we present the streamfunction of the averaged flow field. The streamfunction of the shallow hot Jupiter simulation was presented in \citet{Mendonça2016THOR}. The structure of the Hadley cells is similar, but the relative strength of the Ferrel cells is much stronger than the previous results. In \citet{Mendonça2016THOR}, an additional convection cell structure is present at $\sigma>0.9$. This structure is not observed in our model.

In Figure \ref{fig:stats}, the conservation of physical quantities clearly shows the long spin-up of the model. The time taken for the model to reach equilibrium is about 3000 days, an order of magnitude longer compared to the Held-Suarez test. As expected for a finite volume method, the total mass conserves well, and no change is observed up to the numerical precision. During the simulation, the total AAM and total energy rise by 34.1\% and 1.9\%, respectively.

\section{Conclusion}
In this paper, we introduced a cutting-edge Riemann-Solver-based Finite-Volume Cubed-Sphere dynamic core, ExoCubed. We presented the details and validated the model with various tests ranging from shallow-water tests to general circulation models. We validated the accuracy of the model with standard tests, with or without Coriolis force on a shallow-water sphere. For the general circulation models, we simulated two very different planets, one earth-like (Held-Suarez test) and the other being a hot Jupiter. We provide a variety of Riemann solvers that adeptly manage large temperature/pressure gradients, shock waves, and refine the depiction of rapid atmospheric transitions and high-velocity wind events. The shallow hot Jupiter tests have validated that our model has the capability of dealing with flows in subsonic/sonic regime. Our model can be seamlessly applied to other planetary atmosphere simulations, varying from thin atmospheres such as Mars, Titan, Pluto, super earths, and lava planets, to extended atmospheres relevant to Jupiter, Saturn, Uranus, Neptune, extrasolar gas giants, and brown dwarfs, where projection to an oblate spheroid may be needed (Appendix \ref{sec:oblate}).

The ExoCubed dynamic core employs the Arakawa A-grid \citep{Arakawa1977Computational}, which is an unstaggered grid where all variables are defined at the center of a finite volume. There was a misconception that Arakawa C-grid yielded better phase-propagation properties than Arakawa A-grid. The above statement is only true when the numerical scheme is second-order and when the flow is smooth \citep{harris2021scientific}. \cite{xu2021properties} showed that the phase error of shallow-water waves in unstaggered grids (Arakawa-A) is greatly reduced or eliminated for higher-order numerics. Moreover, when the flow is discontinuous, the C-grid staggering produces far more numerical noise at a discontinuity than an unstaggered grid \citep{chen2018towards}. 

We use the finite volume method to solve the hydrodynamic equations. The design of finite volume methods is based on the control volume and fluxes, which has superior performance in the conservation of the physical quantities, particularly the variables directly solved in the governing equations, such as mass, momentum, and energy. For this reason, the finite volume method is widely used by many existing GCMS (e.g. GFDL FV3, MITgcm). However, unlike the prevailing GCMs, we pioneer the use of Riemann Solvers to solve for fluxes across the cell interface. Using the projection and de-projection method, the ExoCubed dynamic core is agnostic with the selection of the Riemann Solver. For example, the Roe solver \citep{Roe1981Approximate}, the Harten-Lax-van Leer-Einfeldt (HLLE) solver \citep{Harten_Lax_Leer_1983}, and its variation HLLC solver \citep{Toro_Spruce_Speares_1994} are all viable choices. The Riemman Solver applies minimum numerical viscosity to stabilize the numerical solution but is computationally more expensive than traditional methods such as using hyper-viscosity. For pure hydrodynamic problems, the ExoCubed core is slower than other GCMs using hyper-viscosity to dissipate grid-scale noise. Yet, in many cases, the latency is dominated by the computation of physical modules rather than the hydrodynamics itself. So in a full-fledged GCM with comprehensive physics, radiation, and chemistry modules, the ExoCubed-based GCM is expected to perform similarly in terms of speed compared to other models.

To tessellate a sphere, we choose to use the gnomonic equiangular cubed-sphere geometry. This allows a seamless integration with an existing Cartesian geometry based hydrodynamic model. Other options have been explored in the literature, such as using triangular and hexagonal grids (e.g. \citet{Mendonça2016THOR}). The quadrilateral formulation employed here has the advantage of being simple and fast for solving problems in simple geometry. With the transformation of grids to a gnomonic equiangular cubed-sphere, each panel on a sphere is essentially rectangular, and the calculations and parallelization become almost the same as those in Cartesian geometry (see equation \ref{eqn:final}). Utilizing the quadrilateral formulation simplifies the implementation of high-order reconstruction methods. This is achieved by extending the width of the stencil, in consonance with the decision to use the Arakawa A-grid.

Overall, the construction of the ExoCubed dynamic core serves as a general method for extending a Cartesian model to the spherical geometry. The test cases are performed using the SNAP model as the base model \citep{Li20219SNAP} with the VIC scheme \citep{ge2020global} in the vertical. But any base model that solves hydrodynamics using the finite volume method on the structure grid should be able to use the method described in the article. We find that some simple test cases, such as the splash test and the Rossby-Haurwitz wave test are extremely useful for finding errors in the model during the adaptation process.

The code developed in this article has been made publicly available on GitHub at: \url{https://github.com/cshsgy/ExoCubed}.

\section*{Acknowledgement}

C.L. is sponsored by the Heising Simons Foundation under grant AWD023292 and NASA under grant 80NSSC23K0790. We thank Xi Zhang for motivating this project and Huazhi Ge for providing insightful discussions.

\appendix

\section{Conversion to latitudes and longitudes on a sphere}
\subsection{Coordinate transformation}
In practice, for visualizing the results and matching with data, we usually work with latitude-longitude coordinates, which are easier for humans to understand. Here we list the method we use to transform between the gnomonic equiangle coordinates and the lat-lon coordinates.

In panel 1, with a fixed $x_1$, the coordinate line of a constant $x_3=0$ goes along a longitude circle. We define the part where $x_2>0$ as longitude $0^{\circ}$. We denote latitude using $\phi$ and longitude using $\lambda$.

For panel 1, 
\begin{equation}
    \phi=\frac{\pi}{2}-\cos^{-1}\left(\frac{1}{\delta}\right)
\end{equation}
\begin{equation}
    \lambda=\mathrm{atan2}(y,x),
\end{equation}
Here $\textbf{atan2}$ is the 2-argument arctangent function.

For panels 2, 3, 4, and 6,
\begin{equation}
    \phi=\tan^{-1}\left(-\frac{x}{\sqrt{1+(y)^2}}\right)
\end{equation}
\begin{equation}
    \lambda=\lambda_0+\mathrm{atan2}(y,1).
\end{equation}
Here $\phi_0$ are longitudes of the equatorial panels, and the values are $0$,$-\frac{\pi}{2}$, $\frac{\pi}{2}$, and $\pi$ for panels 2, 3, 4, and 6, respectively.

For panel 5,
\begin{equation}
    \phi=-\frac{\pi}{2}+\cos^{-1}\left(\frac{1}{\delta}\right)
\end{equation}
\begin{equation}
    \lambda=\mathrm{atan2}(y,-x)
\end{equation}

\subsection{Vector transformation}
In addition to the conversion of local coordinates, we also need to project the velocity vectors from the gnomonic equiangle coordinates $(V_x,V_y)$ to the local cartesian that goes along the zonal and meridional directions $(V_\lambda,V_\phi)$, or $(U,V)$. The basis vectors for the lat-lon grid are
\begin{align}
\hat{\boldsymbol{\phi}} &= \begin{pmatrix}
-\sin\phi \cos\lambda \\
-\sin\phi \sin\lambda \\
\cos\phi
\end{pmatrix} \\
\hat{\boldsymbol{\lambda}} &= \begin{pmatrix}
-\sin\lambda \\
\cos\lambda \\
0
\end{pmatrix},
\end{align}
The equatorial panels are symmetric, so they will share the same way for vector decomposition. Writing the unit basis vectors in the local cartesian coordinates,
\begin{equation}
    \hat{\boldsymbol{\phi}}=\frac{1}{D\delta}(x,-D^2,xy)
\end{equation}
\begin{equation}
    \hat{\boldsymbol{\lambda}}=\frac{1}{D}(-y,0,1),
\end{equation}
The inner product is obtained with the gnomonic equiangle coordinate basis
\begin{equation}
\hat{\boldsymbol{\phi}}\cdot{\textbf{b}}_2=-1,\;\hat{\boldsymbol{\lambda}}\cdot{\textbf{b}}_2=0
\end{equation}
\begin{equation}
\hat{\boldsymbol{\phi}}\cdot{\textbf{b}}_3=\frac{xy}{CD},\; \hat{\boldsymbol{\lambda}}\cdot{\textbf{b}}_3=\frac{\delta}{CD},
\end{equation}
Thus the transformation matrix can be written as
\begin{equation}
    \begin{pmatrix}
U\\
V
\end{pmatrix}
=
\begin{pmatrix}
    0 & \frac{\delta}{CD} \\
    -1 & \frac{xy}{CD}
\end{pmatrix}
\begin{pmatrix}
V_x\\
V_y
\end{pmatrix}
\end{equation}

\begin{equation}
    \begin{pmatrix}
V_x\\
V_y
\end{pmatrix}
=
\begin{pmatrix}
    \frac{xy}{\delta} & -1 \\
    \frac{CD}{\delta} & 0
\end{pmatrix}
\begin{pmatrix}
U\\
V
\end{pmatrix}.
\end{equation}

For panel 1, for the simplicity of the expressions, we define
\begin{equation}
    E=\sqrt{x^2+y^2},
\end{equation}
So that
\begin{equation}
    \hat{\boldsymbol{\phi}}=\frac{1}{E\delta}(E^2,-x,-y)
\end{equation}
\begin{equation}
    \hat{\boldsymbol{\lambda}}=\frac{1}{E}(0,-y,x),
\end{equation}

\begin{equation}
\hat{\boldsymbol{\phi}}\cdot{\textbf{b}}_2=-\frac{x}{DE},\;\hat{\boldsymbol{\lambda}}\cdot{\textbf{b}}_2=-\frac{y}{CE}
\end{equation}
\begin{equation}
\hat{\boldsymbol{\phi}}\cdot{\textbf{b}}_3=-\frac{\delta y}{DE},\;\hat{\boldsymbol{\lambda}}\cdot{\textbf{b}}_3=\frac{\delta x}{CE},
\end{equation}
Thus the transformation matrix can be written as
\begin{equation}
    \begin{pmatrix}
U\\
V
\end{pmatrix}
=
\begin{pmatrix}
    -\frac{\delta y}{DE} & \frac{\delta x}{CE} \\
    -\frac{x}{DE} & -\frac{y}{CE}
\end{pmatrix}
\begin{pmatrix}
V_x\\
V_y
\end{pmatrix}
\end{equation}

\begin{equation}
    \begin{pmatrix}
V_x\\
V_y
\end{pmatrix}
=
\begin{pmatrix}
    -\frac{Dy}{\delta E} & -\frac{Dx}{E} \\
    \frac{Cx}{\delta E} & -\frac{Cy}{E}
\end{pmatrix}
\begin{pmatrix}
U\\
V
\end{pmatrix}.
\end{equation}
The transformation for panel 5 can be calculated in the same way except for the inversion of the sign
\begin{equation}
    \begin{pmatrix}
U\\
V
\end{pmatrix}
=
\begin{pmatrix}
    \frac{\delta y}{DE} & -\frac{\delta x}{CE} \\
    \frac{x}{DE} & \frac{y}{CE}
\end{pmatrix}
\begin{pmatrix}
V_x\\
V_y
\end{pmatrix}
\end{equation}

\begin{equation}
    \begin{pmatrix}
V_x\\
V_y
\end{pmatrix}
=
\begin{pmatrix}
    \frac{Dy}{\delta E} & \frac{Dx}{E} \\
    -\frac{Cx}{\delta E} & \frac{Cy}{E}
\end{pmatrix}
\begin{pmatrix}
U\\
V
\end{pmatrix}.
\end{equation}

\section{Coriolis force}
The Coriolis acceleration differs depending on the cubed-sphere panel. In the Cartesian coordinate with $(z,x,y)$ ordering, the planetary vorticity is $2\mathbf{\Omega}=(2\Omega,0,0)$. In a matrix form, the Coriolis acceleration for the polar panels is:
\begin{equation}
    \Psi_{P} = \frac{2\Omega CD}{(\delta)^2} \begin{bmatrix}
        0 & yC & -xD \\
        -yC & 0 & -1 \\
        xD & 1 & 0,
    \end{bmatrix}
\end{equation}
and for the equatorial panels is:
\begin{equation}
    \Psi_{E} = \frac{2\Omega CD}{(\delta)^2} \begin{bmatrix}
        0 & 0 & -D\\
        0 & 0 & x \\
        D & -x & 0
    \end{bmatrix}
\end{equation}
\section{Cubed Oblate Spheroid}
\label{sec:oblate}
For problems related to giant planets and brown dwarfs, it may be useful to use the projection onto an oblate spheroid instead of a sphere. In a spherical geometry, the vertical (radial) direction is special because it is normal to the other two horizontal directions. Thus, the covariant and the contra-variant basis vectors, $\mathbf{b}_1$ and $\mathbf{b}^1$ are the same. In an oblate spheroid, it is crucial to distinguish between the ``vertical'' direction, which is normal to a geopotential surface, and the ``radial'' direction. The ``vertical'' direction is along the contra-variant basis vector, $\mathbf{b}^1$ and the ``radial'' direction is along the covariant basis vector, $\mathbf{b}_1$. This definition is consistent with the definition of basis vectors in the spherical geometry (Equations \ref{eqn:b1} - \ref{eqn:b3}). The methods described in the main article can be easily extended to an oblate spheroid by generalizing the relation between the radial direction and two horizontal directions. We outline the generalization in the following paragraphs.

For an oblate spheroid with eccentricity $e$, equation (\ref{eqn:delta}) and the following derivations take a different form. Here we consider a ``unit spheroid'' in the Cartesian coordinate $(z,x,y)$:
\begin{equation}
    (x)^2+(y)^2+\frac{(z)^2}{1-e^2}=1
\end{equation}
Then the position vector of a point in the spherical panel is
\begin{equation}
    \mathbf{r}=\frac{r}{\delta(n)}\left(1,x,y\right)
\end{equation}
Where $x$ and $y$ are as previously defined $x=\tan \xi$ and $y=\tan\eta$. Due to the loss of symmetry in an oblate spheroid, the values of $\delta(n)$ depend on the panel number $n$. In polar panels (top and bottom),
\begin{equation}
    \delta(1,5)=\sqrt{(x)^2+(y)^2+\frac{1}{1-e^2}},
\end{equation}
In the left and right equatorial panels,
\begin{equation}
    \delta(3,4)=\sqrt{\frac{(x)^2}{1-e^2}+(y)^2+1},
\end{equation}
In the front and back equatorial panels,
\begin{equation}
    \delta(2,6)=\sqrt{(x)^2+\frac{(y)^2}{1-e^2}+1}.
\end{equation}

With the formula of the position vector determined, following the same line as Equations (\ref{eqn:delta})-(\ref{eqn:b3}), the basis vectors can be determined. Subsequently, the metric tensors and the Christoffel symbols can be found in Equations (\ref{eq:mettensor})-(\ref{eqn:gamma2}). 

It is worth noting that the direction of the effective gravitational acceleration is along the ``vertical'' direction rather than the ``radial'' direction in an oblate spheroid geometry. So, effective gravitational acceleration has only one component in the contra-variant basis vectors and has all three components in the covariant basis vectors, i.e.:

\begin{equation}
    \nabla\phi = g_1 \mathbf{b}^1 = g^1 \mathbf{b}_1 + g^2 \mathbf{b}_2 + g^3 \mathbf{b}_3 \label{eqn:g3}
\end{equation}

Equation (\ref{eqn:g3}) reveals an additional advantage of solving for covariant momentum $(\rho v_\alpha)$ instead of contra-variant momentum $(\rho v^\alpha)$: in a covariant formulation, all three momentum equations remain unchanged, with gravitational acceleration having no impact on the horizontal covariant momenta, even in the context of oblate spheroid geometry. Consequently, the equations of motion, Equations (\ref{eq:mom})-(\ref{eqn:final}), remain unchanged.

Next, in the projection step within the Riemann Solver, the basis vectors $\mathbf{b}_1$ and $\mathbf{b}^1$ are no longer equivalent. 
Hence, projections are needed for all surfaces in three directions. 
The projection and deprojection relationships can be inferred in the same way as Equations (\ref{eq:e1b1})-(\ref{eq:O3-1}), projecting to a local coordinate basis formed by the normalized contra-variant basis normal to this surface and other two covariant bases. 
After the projection, the Riemann Solver step described in Equations (\ref{eq:riemann1})-(\ref{eq:riemann4}) also remains the same.

Despite that the oblate spheroid has a more complex geometry, the convenience that only a one-dimensional interpolation is needed for ghost zones is preserved. The formula and intuition provided by Equation (\ref{eqn:interp}) and Figure \ref{fig:in-panel-sync} remain valid. The transformation of vectors across the panel boundary can be similarly derived, as $\mathbf{b}_1$ is unchanged when a panel boundary is crossed, so the transformation is still from $\{\mathbf{b}_1,\mathbf{b}_2,\mathbf{b}_3\}$ to $\{\mathbf{b}_1,\mathbf{b}_2',\mathbf{b}_3'\}$. Finally, the metric terms can be calculated according to Equation (\ref{eqn:ma}), with more complex Christoffel symbols calculated for the oblate spheroid coordinate basis.

One final generalization involves the ``vertical" implicit solver. In the spherical geometry, the ``vertical" velocity is identical in both its covariant and contra-variant forms. However, in an oblate spheroid, the velocity differs between the ``vertical" direction (contravariant component) and the ``radial" direction (covariant component). The original vertical implicit solver was developed for a spherical-polar grid. New implicit solver should be developed using covariant momentum in an oblate spheroid.

\bibliography{main}{}

\begin{thebibliography}{}
\expandafter\ifx\csname natexlab\endcsname\relax\def\natexlab#1{#1}\fi
\providecommand{\url}[1]{\href{#1}{#1}}
\providecommand{\dodoi}[1]{doi:~\href{http://doi.org/#1}{\nolinkurl{#1}}}
\providecommand{\doeprint}[1]{\href{http://ascl.net/#1}{\nolinkurl{http://ascl.net/#1}}}
\providecommand{\doarXiv}[1]{\href{https://arxiv.org/abs/#1}{\nolinkurl{https://arxiv.org/abs/#1}}}

\bibitem[{Adcroft {et~al.}(2004{\natexlab{a}})Adcroft, Campin, Hill, \&
  Marshall}]{adcroft2004implementation}
Adcroft, A., Campin, J.-M., Hill, C., \& Marshall, J. 2004{\natexlab{a}},
  Monthly Weather Review, 132, 2845

\bibitem[{Adcroft {et~al.}(2004{\natexlab{b}})Adcroft, Hill, Campin, Marshall,
  \& Heimbach}]{adcroft2004overview}
Adcroft, A., Hill, C., Campin, J.-M., Marshall, J., \& Heimbach, P.
  2004{\natexlab{b}}, in Proceedings of the ECMWF seminar series on Numerical
  Methods, Recent developments in numerical methods for atmosphere and ocean
  modelling, 139--149

\bibitem[{Arakawa \& Lamb(1977)}]{Arakawa1977Computational}
Arakawa, A., \& Lamb, V.~R. 1977, Methods in Computational Physics: Advances in
  Research and Applications, 173–265,
  \dodoi{10.1016/b978-0-12-460817-7.50009-4}

\bibitem[{Bacon {et~al.}(2000)Bacon, Ahmad, Boybeyi, Dunn, Hall, Lee, Sarma,
  Turner, Waight, Young, \& Zack}]{Bacon2000OMEGA}
Bacon, D.~P., Ahmad, N.~N., Boybeyi, Z., {et~al.} 2000, Monthly Weather Review,
  128, 2044 , \dodoi{10.1175/1520-0493(2000)128<2044:ADAWAD>2.0.CO;2}

\bibitem[{Bell \& Cowan(2018)}]{Bell2018Increased}
Bell, T.~J., \& Cowan, N.~B. 2018, The Astrophysical Journal Letters, 857,
  \dodoi{10.3847/2041-8213/aabcc8}

\bibitem[{Bindle {et~al.}(2021)Bindle, Martin, Cooper, Lundgren, Eastham, Auer,
  Clune, Weng, Lin, Murray, Meng, Keller, Putman, Pawson, \&
  Jacob}]{Bindle2021Stretch}
Bindle, L., Martin, R.~V., Cooper, M.~J., {et~al.} 2021, Geoscientific Model
  Development, 14, 5977, \dodoi{10.5194/gmd-14-5977-2021}

\bibitem[{Bretherton(1969)}]{Bretherton1969lamb}
Bretherton, F.~P. 1969, Quarterly Journal of the Royal Meteorological Society,
  95, 754, \dodoi{https://doi.org/10.1002/qj.49709540608}

\bibitem[{Chen(2021)}]{chen2021lmars}
Chen, X. 2021, Journal of Advances in Modeling Earth Systems, 13, e2020MS002280

\bibitem[{Chen {et~al.}(2018)Chen, Lin, \& Harris}]{chen2018towards}
Chen, X., Lin, S.-J., \& Harris, L.~M. 2018, Journal of Advances in Modeling
  Earth Systems, 10, 2333

\bibitem[{Christie {et~al.}(2022)Christie, Lee, Innes, Noti, Charnay, Fauchez,
  \& et~al.}]{DuncanArXivCamembert}
Christie, D.~A., Lee, E. K.~H., Innes, H., {et~al.} 2022, Camembert: A
  Mini-Neptunes GCM Intercomparison, protocol version 1.0. A cuisines model
  Intercomparison project.
\newblock \url{https://ar5iv.labs.arxiv.org/html/2211.04048v1}

\bibitem[{Constantinou \& Madhusudhan(2022)}]{constantinou2022characterizing}
Constantinou, S., \& Madhusudhan, N. 2022, Monthly Notices of the Royal
  Astronomical Society, 514, 2073

\bibitem[{Deitrick {et~al.}(2020)Deitrick, Mendon{\c c}a, Schroffenegger,
  Grimm, Tsai, \& Heng}]{deitrick2020thor2}
Deitrick, R., Mendon{\c c}a, J., Schroffenegger, U., {et~al.} 2020,
  Astrophysical Journal Supplement Series, 248,
  \dodoi{10.3847/1538-4365/ab930e}

\bibitem[{Dowling {et~al.}(1998)Dowling, Fischer, Gierasch, Harrington, LeBeau,
  \& Santori}]{Dowling1998epic}
Dowling, T., Fischer, A., Gierasch, P., {et~al.} 1998, Icarus, 132, 221–238,
  \dodoi{10.1006/icar.1998.5917}

\bibitem[{Fromang {et~al.}(2016)Fromang, Leconte, \& Heng}]{fromang2016shear}
Fromang, S., Leconte, J., \& Heng, K. 2016, Astronomy \& Astrophysics, 591,
  A144

\bibitem[{Gammie {et~al.}(2003)Gammie, McKinney, \& T{\'o}th}]{gammie2003harm}
Gammie, C.~F., McKinney, J.~C., \& T{\'o}th, G. 2003, The Astrophysical
  Journal, 589, 444

\bibitem[{Ge {et~al.}(2020)Ge, Li, Zhang, \& Lee}]{ge2020global}
Ge, H., Li, C., Zhang, X., \& Lee, D. 2020, The Astrophysical Journal, 898, 130

\bibitem[{Godunov \& Bohachevsky(1959)}]{Godunov1959Finite}
Godunov, S.~K., \& Bohachevsky, I. 1959, {Matemati{\v c}eskij sbornik}, 47(89),
  271.
\newblock \url{https://hal.science/hal-01620642}

\bibitem[{Grinfeld(2014)}]{Grinfeld2014Introduction}
Grinfeld, P. 2014, Introduction to Tensor Analysis and the Calculus of Moving
  Surfaces (New York: Springer)

\bibitem[{Harris {et~al.}(2021)Harris, Chen, Putman, Zhou, \&
  Chen}]{harris2021scientific}
Harris, L., Chen, X., Putman, W., Zhou, L., \& Chen, J.-H. 2021.
\newblock \url{https://repository.library.noaa.gov/view/noaa/30725}

\bibitem[{Harris {et~al.}(2020)Harris, Chen, Zhou, \&
  Chen}]{harris2020nonhydrostatic}
Harris, L., Chen, X., Zhou, L., \& Chen, J.-H. 2020

\bibitem[{Harten {et~al.}(1983)Harten, Lax, \& Leer}]{Harten_Lax_Leer_1983}
Harten, A., Lax, P.~D., \& Leer, B.~v. 1983, SIAM Review, 25, 35–61,
  \dodoi{10.1137/1025002}

\bibitem[{Held \& Suarez(1994)}]{HS94}
Held, I.~M., \& Suarez, M.~J. 1994, Bulletin of the American Meteorological
  Society, 75, 1825 ,
  \dodoi{https://doi.org/10.1175/1520-0477(1994)075<1825:APFTIO>2.0.CO;2}

\bibitem[{Heng {et~al.}(2011)Heng, Menou, \& Phillipps}]{Heng2011Atmos}
Heng, K., Menou, K., \& Phillipps, P.~J. 2011, Monthly Notices of the Royal
  Astronomical Society, 413, 2380, \dodoi{10.1111/j.1365-2966.2011.18315.x}

\bibitem[{Kaspi \& Showman(2015)}]{kaspi2015atmospheric}
Kaspi, Y., \& Showman, A.~P. 2015, The Astrophysical Journal, 804, 60

\bibitem[{Komacek {et~al.}(2021)Komacek, Kang, Lustig-Yaeger, \&
  Olson}]{komacek2021constraining}
Komacek, T.~D., Kang, W., Lustig-Yaeger, J., \& Olson, S.~L. 2021, Elements,
  17, 251, \dodoi{10.2138/gselements.17.4.251}

\bibitem[{Komacek \& Showman(2016)}]{Komacek2016Atmospheric}
Komacek, T.~D., \& Showman, A.~P. 2016, The Astrophysical Journal, 821, 16,
  \dodoi{10.3847/0004-637x/821/1/16}

\bibitem[{Kuma {et~al.}(2023)Kuma, Bender, \& J{\"o}nsson}]{kuma2023climate}
Kuma, P., Bender, F. A.-M., \& J{\"o}nsson, A.~R. 2023, Journal of Advances in
  Modeling Earth Systems, 15, e2022MS003588

\bibitem[{LeVeque(2002)}]{leveque2002finite}
LeVeque, R.~J. 2002, Finite volume methods for hyperbolic problems, Vol.~31
  (Cambridge university press)

\bibitem[{Li \& Chen(2019)}]{Li20219SNAP}
Li, C., \& Chen, X. 2019, The Astrophysical Journal

\bibitem[{Lin(1997)}]{lin1997finite}
Lin, S.-J. 1997, Quarterly Journal of the Royal Meteorological Society, 123,
  1749

\bibitem[{Lin(2004)}]{lin2004vertically}
---. 2004, Monthly Weather Review, 132, 2293

\bibitem[{Mendonça {et~al.}(2016)Mendonça, Grimm, Grosheintz, \&
  Heng}]{Mendonça2016THOR}
Mendonça, J.~M., Grimm, S.~L., Grosheintz, L., \& Heng, K. 2016, The
  Astrophysical Journal, 829, 115, \dodoi{10.3847/0004-637X/829/2/115}

\bibitem[{Menou \& Rauscher(2009)}]{Menou2009Atmos}
Menou, K., \& Rauscher, E. 2009, The Astrophysical Journal

\bibitem[{Mouallem {et~al.}(2023)Mouallem, Harris, \&
  Chen}]{mouallem2023implementation}
Mouallem, J., Harris, L., \& Chen, X. 2023, Journal of Advances in Modeling
  Earth Systems, 15, e2023MS003712

\bibitem[{Nair {et~al.}(2005)Nair, Thomas, \& Loft}]{nair2005discontinuous}
Nair, R.~D., Thomas, S.~J., \& Loft, R.~D. 2005, Monthly weather review, 133,
  876

\bibitem[{Nasr {et~al.}(2022)Nasr, Bradley, Lewis, Hollingsworth, \&
  Dowling}]{Nasr_2022}
Nasr, C.-R.~C., Bradley, M.~E., Lewis, S.~R., Hollingsworth, J.~L., \& Dowling,
  T.~E. 2022, The Planetary Science Journal, 3, 165, \dodoi{10.3847/PSJ/ac72ab}

\bibitem[{Neale(2010)}]{Neale2010CAM}
Neale, R.~B. 2010.
\newblock
  \url{https://www2.cesm.ucar.edu/models/ccsm4.0/cam/docs/description/cam4_desc.pdf}

\bibitem[{Pedlosky(2013)}]{pedlosky2013geophysical}
Pedlosky, J. 2013, Geophysical fluid dynamics (Springer Science \& Business
  Media)

\bibitem[{Putman \& Lin(2007)}]{putman2007finite}
Putman, W.~M., \& Lin, S.-J. 2007, Journal of Computational Physics, 227, 55

\bibitem[{Robert(1993)}]{Robert1993Bubble}
Robert, A. 1993, Journal of the Atmospheric Sciences, 50, 1865–1873,
  \dodoi{10.1175/1520-0469(1993)050&lt;1865:bcewas&gt;2.0.co;2}

\bibitem[{Roe(1981)}]{Roe1981Approximate}
Roe, P. 1981, Journal of Computational Physics, 43, 357–372,
  \dodoi{10.1016/0021-9991(81)90128-5}

\bibitem[{Ronchi {et~al.}(1996)Ronchi, Iacono, \& Paolucci}]{ronchi1996cubed}
Ronchi, C., Iacono, R., \& Paolucci, P.~S. 1996, Journal of computational
  physics, 124, 93

\bibitem[{Sadourny(1972)}]{sadourny1972conservative}
Sadourny, R. 1972, Monthly Weather Review, 100, 136

\bibitem[{Sadourny {et~al.}(1968)Sadourny, Arakawa, \&
  Mintz}]{sadourny1968integration}
Sadourny, R., Arakawa, A., \& Mintz, Y. 1968, Monthly Weather Review, 96, 351

\bibitem[{Showman \& Guillot(2002)}]{showman2002atmospheric}
Showman, A.~P., \& Guillot, T. 2002, Astronomy \& Astrophysics, 385, 166

\bibitem[{Showman {et~al.}(2015)Showman, Lewis, \& Fortney}]{Showman2015three}
Showman, A.~P., Lewis, N.~K., \& Fortney, J.~J. 2015, The Astrophysical Journal

\bibitem[{Smith \& Dritschel(2006)}]{Smith2006Revisiting}
Smith, R.~K., \& Dritschel, D.~G. 2006, Journal of Computational Physics, 217,
  473, \dodoi{https://doi.org/10.1016/j.jcp.2006.01.011}

\bibitem[{Stone {et~al.}(2020)Stone, Tomida, White, \&
  Felker}]{stone2020athena++}
Stone, J.~M., Tomida, K., White, C.~J., \& Felker, K.~G. 2020, The
  Astrophysical Journal Supplement Series, 249, 4

\bibitem[{Straka {et~al.}(1993)Straka, Wilhelmson, Wicker, Anderson, \&
  Droegemeier}]{Straka1993Numerical}
Straka, J.~M., Wilhelmson, R.~B., Wicker, L.~J., Anderson, J.~R., \&
  Droegemeier, K.~K. 1993, International Journal for Numerical Methods in
  Fluids, 17, 1–22, \dodoi{10.1002/fld.1650170103}

\bibitem[{Thuburn \& Li(2000)}]{Thuburn2000Numerical}
Thuburn, J., \& Li, Y. 2000, Tellus, 52A, 181

\bibitem[{Toro {et~al.}(1994)Toro, Spruce, \&
  Speares}]{Toro_Spruce_Speares_1994}
Toro, E.~F., Spruce, M., \& Speares, W. 1994, Shock Waves, 4, 25–34,
  \dodoi{10.1007/bf01414629}

\bibitem[{Ullrich \& Jablonowski(2012)}]{Ullrich2012MCore}
Ullrich, P.~A., \& Jablonowski, C. 2012, Journal of Computational Physics, 231,
  5078, \dodoi{https://doi.org/10.1016/j.jcp.2012.04.024}

\bibitem[{Ullrich {et~al.}(2010)Ullrich, Jablonowski, \&
  Van~Leer}]{ullrich2010high}
Ullrich, P.~A., Jablonowski, C., \& Van~Leer, B. 2010, Journal of Computational
  Physics, 229, 6104

\bibitem[{White {et~al.}(2016)White, Stone, \& Gammie}]{white2016extension}
White, C.~J., Stone, J.~M., \& Gammie, C.~F. 2016, The Astrophysical Journal
  Supplement Series, 225, 22

\bibitem[{Williamson {et~al.}(1992)Williamson, Drake, Hack, Jakob, \&
  Swarztrauber}]{williamson1992standard}
Williamson, D.~L., Drake, J.~B., Hack, J.~J., Jakob, R., \& Swarztrauber, N.
  1992, Journal of Computational Physics, 102, 211

\bibitem[{Xu {et~al.}(2021)Xu, Chen, \& Wu}]{xu2021properties}
Xu, D., Chen, D., \& Wu, K. 2021, Advances in Atmospheric Sciences, 38, 615

\end{thebibliography}
\bibliographystyle{aasjournal}

\end{document}